\documentclass[11pt,a4paper]{article}
\usepackage[margin=2.5cm]{geometry}
\usepackage{authblk}
\usepackage{graphicx}
\usepackage{booktabs}
\usepackage{anyfontsize}
\usepackage{colortbl}
\usepackage{xcolor}
\usepackage{amsmath,amssymb,amsthm}
\usepackage{bm}
\usepackage{algorithm}
\usepackage{comment}
\usepackage{algpseudocode}
\usepackage[numbers,sort&compress]{natbib}
\usepackage{hyperref}
\graphicspath{{./figures/}}

\theoremstyle{remark}
\newtheorem{remark}{Remark}

\newcommand{\AM}[1]{\textcolor{black}{#1}}
\newcommand{\LL}[1]{\textcolor{black}{#1}}

\newcommand{\bx}{\mathbf{x}}
\newcommand{\bu}{\mathbf{u}}
\newcommand{\bv}{\mathbf{v}}
\newcommand{\bn}{\mathbf{n}}
\newcommand{\bg}{\mathbf{g}}
\newcommand{\bsig}{\boldsymbol{\sigma}}
\newcommand{\beps}{\boldsymbol{\varepsilon}}
\newcommand{\bI}{\mathbf{I}}
\newcommand{\Cel}{\mathbb{C}}
\newcommand{\gw}{\gamma_w}
\newcommand{\q}{\mathbf{q}}
\newcommand{\tr}{\operatorname{tr}}

\title{A hyper-reduction strategy for elastoplastic
hydro-mechanical problems in partially saturated soil}

\author[1,2]{Alba Muix\'i}
\author[2]{Llu\'is Monforte}
\author[1]{Antonio Rodr\'iguez-Ferran}
\author[1,2]{Sergio Zlotnik\thanks{Corresponding author: \texttt{sergio.zlotnik@upc.edu}}}

\affil[1]{Laboratori de C\`alcul Num\`eric (LaC\`aN),
Universitat Polit\`ecnica de Catalunya - BarcelonaTech (UPC),
Barcelona, Spain}
\affil[2]{Centre Internacional de M\`etodes Num\`erics en
Enginyeria (CIMNE),
Barcelona, Spain}

\date{}

\begin{document}

\maketitle

\begin{abstract}
	This work proposes a hyper-reduced order model for coupled hydro-mechanical simulations of elastoplastic deformation and water flow in partially saturated soils. These simulations are used to monitor and assess the safety of geotechnical structures such as earthfill dams, where calibration, uncertainty quantification and scenario analysis require solving the same transient model many times for different parameter values.
	The displacement and the pore pressure are approximated in a field-split reduced basis obtained by proper orthogonal decomposition, and the Galerkin-projected equations are integrated with decomposed cubature rules, one per balance equation, built with the energy-conserving sampling and weighting (ECSW) method. 
	Each residual is split into an affine part and a nonlinear correction: the affine part is computed exactly, at negligible online cost, so that the cubature rule only needs to approximate the correction, which is localized in space. 
	The construction is independent of the constitutive model choices and is implemented in FEniCSx by assembling the original variational forms on the small set of elements selected by ECSW.
	The method is tested on two- and three-dimensional dams retaining tailings and water, simulating a raise of the dam and the subsequent consolidation phase, with the friction angle, stiffness and hydraulic conductivity of the dam material and the height of the raise as varying parameters. 
	The hyper-reduced model reproduces the displacement and pressure fields with errors below $0.5\%$ with respect to the finite element solution, achieving speed-ups of up to $43\times$ in two dimensions and $77\times$ in three. 
	These performance gains grow with the extent of the plastic yielding, and the selected elements concentrate in the plastified and unsaturated zones, whose size is set by the physics rather than by the mesh resolution.
\end{abstract}


%========================================================================

\section{Introduction}\label{sec:intro}

Numerical models are increasingly used together with sensor measurements to monitor and manage large geostructures such as earthfill dams \cite{Salazar2017}. Data assimilation keeps these models calibrated against field observations, and the updated models then support decision making, optimization, control and uncertainty quantification, ultimately enabling digital twins that evolve with the structure throughout its service life \cite{Mohsan2023,Ding2025}. These applications are inherently many-query problems: the same numerical model must be evaluated repeatedly for different values of its parameters and, to support decisions during operation, these evaluations must be in real time.

In the case of earthfill dams, the model of interest is a coupled hydro-mechanical system, in which pore water flow and deformation of the soil skeleton interact bidirectionally \cite{LewisSchrefler1998,ZienkiewiczCompGeo1999}. Water contributes to the self-weight of the soil and influences its strength through the effective stress, while deformation of the solid skeleton modifies the pore space and drives the flow. For partially saturated soils, a realistic description combines unsaturated flow with an elastoplastic model of the skeleton. Each model evaluation then requires the solution of a nonlinear coupled finite element problem over hundreds of time steps, and repeated evaluations quickly become computationally prohibitive.

Projection-based Reduced-Order Models (ROMs) provide an effective framework for accelerating many-query simulations \cite{RozzaRB2008,BennerSurvey2015,QuarteroniRB2016,HesthavenRB2016}. 
In an offline stage, a low-dimensional basis is constructed from a training set of high-fidelity solutions
computed for representative parameter values. Online, the solution for a new value of the parameters is sought in the space spanned by this basis, and the governing equations are enforced through a Galerkin projection, which leads to a reduced system with far fewer unknowns.  
In geomechanics, projection-based ROMs have been applied to groundwater flow \cite{Vermeulen2004,Pasetto2013}, coupled flow--geomechanics problems in reservoir engineering \cite{Jin2020}, and coupled thermo-hydro-mechanical systems \cite{Larion2020,Larion2024}.
For earthfill dams, Nasika et al.~\cite{Nasika2022} applied a ROM based on the Proper Orthogonal Decomposition (POD) to the hydro-mechanical model, and the same ROM was used in \cite{ShiraInverse} to calibrate the model parameters from piezometer measurements.

For nonlinear problems, however, Galerkin projection alone provides only limited computational savings. Although the reduced system is small, the nonlinear residual and Jacobian must still be assembled over the full mesh at every Newton--Raphson iteration, requiring evaluation of the constitutive model at all integration points. Hyper-reduction addresses this bottleneck by approximating the reduced operators and avoiding the assembly of the full-order system \cite{Ryckelynck2005}. Existing approaches can be broadly classified into two families. The first approximates the integration of the reduced operators over a carefully selected subset of elements or quadrature points, as in the Reduced Integration Domain (RID) \cite{Ryckelynck2005,Ryckelynck2009}, the Energy-Conserving Sampling and Weighting (ECSW) method \cite{FarhatECSW2014,FarhatECSW2015,Grimberg2021}, the Empirical Cubature Method (ECM) \cite{ECM2017} and other empirical quadrature procedures \cite{YanoPatera2019}. The second approximates the nonlinear operators themselves using reduced bases, with online coefficients obtained through interpolation at selected entries, as in the Discrete Empirical Interpolation Method (DEIM) \cite{DEIM2010}, matrix DEIM \cite{NegriMDEIM2015}, gappy POD \cite{BuiThanhWillcox2004} and Gauss--Newton with Approximated Tensors (GNAT) \cite{GNAT2013}, or by projection of the nonlinear terms \cite{MuixiScaffolds2022}.

Both families have been extended to coupled and multi-field problems. In geomechanics, POD--DEIM reduced models have been developed for nonlinear groundwater flow \cite{Stanko2016}, and Nasika et al.~\cite{Nasika2023} applied DEIM to the coupled hydro-mechanical problem, constructing separate approximations for the hydraulic and the mechanical residual blocks; a localized variant (LDEIM) further clustered the training data into regimes of behaviour to reduce the size of the approximation bases. The mechanical response remained elastic, so the nonlinear behaviour originated solely from the flow problem. Thermo-hydro-mechanical models for nuclear waste disposal have also been reduced, assuming fully saturated conditions and a linear elastic mechanical response \cite{IolloTHM2022, Larion2020}.  
Larion et al.~\cite{Larion2024} did consider plasticity, but avoided hyper-reduction by confining the constitutive evaluations to a pre-selected subdomain, which must contain the plastic zone for every parameter value. 
In computational solid mechanics, Zhang, Kehls and co-workers proposed field-wise strategies based on DEIM and ECSW for thermo-mechanical models coupled with damage \cite{Zhang2025Decomposed} and for damage-plasticity \cite{KehlsDecomposed2025}, reporting improved accuracy and computational efficiency with ECSW.

In this work, we propose a hyper-reduction strategy for coupled elastoplastic hydro-mechanical problems in partially saturated soils, where strong nonlinearities of a different nature arise simultaneously in both governing equations. 
To the best of our knowledge, hyper-reduction of this class of problems remains unaddressed.

The choice of the specific hyper-reduction approach is conditioned by the implementation framework. Modern finite element libraries such as FEniCSx \cite{fenicsx2023}, Firedrake \cite{Firedrake2016} or FreeFEM \cite{FreeFEM2012} enable a high-level implementation of complex physical models. 
The variational forms are declared in near-mathematical notation and their assembly is automatic, which makes coupled nonlinear problems such as the present one affordable to develop and to maintain. 
Here, the model is implemented in FEniCSx.
Within this setting, however, residuals are available only as assembled forms. Interpolation-based methods, which reconstruct the nonlinear operators from a small number of their entries, would require intrusive modifications of the assembly process, together with a careful treatment of the internal variables of history-dependent constitutive models at the sampled entries. 
Quadrature-based methods modify only the integration rule, and those selecting individual quadrature points \cite{ECM2017,YanoPatera2019} would still require an intrusive modification of assembly routines.
Quadrature methods selecting complete elements \cite{Ryckelynck2005,FarhatECSW2014} are instead directly compatible with standard finite element assembly, which already operates element by element. Among them, ECSW additionally constrains the cubature weights to be non-negative, which preserves the structure of the Galerkin projection and contributes to the robustness of the method \cite{FarhatECSW2014,FarhatECSW2015}. These properties make ECSW the natural candidate for our problem.

The construction of an efficient ECSW rule must take into account several features of the coupled problem. 
The displacement and pore pressure fields exhibit fundamentally different nonlinear behaviour, localized in different regions: plasticity develops in yield zones whose location and extent vary significantly with the material parameters, whereas the hydraulic nonlinearity originates in the retention curves and concentrates near the phreatic surface. 
Also, each residual consists of linear and nonlinear terms; a rule trained on the complete residual therefore spends elements reproducing linear contributions that could be computed exactly at negligible cost.
Finally, the two residual blocks (\LL{mechanical} and flow) may differ in magnitude and in the proportion of their content that is genuinely nonlinear.

To address these challenges, we propose a field-wise formulation of both the reduced bases and the ECSW cubature. 
Independent POD bases are constructed for the displacement and pore pressure fields, resulting in a field-split reduced space. 
The ECSW cubature is then decomposed into two independent rules, one for each governing equation. Rather than approximating the complete residual, each rule is trained on the nonlinear correction with respect to an exact affine baseline, consisting of an elastic stress response for the \LL{mechanical} problem and a mean-state approximation of the retention coefficients for the hydraulic problem. This formulation isolates the localized nonlinear contributions that must be captured by the hyper-reduced model, and further reduces the size of the reduced mesh at equal accuracy.

The proposed methodology does not require modifications specific to the chosen constitutive formulation, requires no additional full-order simulations beyond those used to construct the POD reduced basis, and is implemented in FEniCSx exclusively through standard assembly routines. 
\AM{The strategy is assessed on numerical experiments simulating two- and three-dimensional earthfill dams that retain tailings and water, through a raise of the dam and the subsequent consolidation \LL{phase}. Four parameters vary across the simulations, namely the friction angle $\phi$, the Young's modulus $E$, the saturated conductivity $K_s$ and the height of the raise $h_{\mathrm{fill}}$. In these experiments, the hyper-reduced solutions remain within the accuracy range of the reduced basis model while achieving speed-ups between $7\times$ and $77\times$.} Moreover, the reduced meshes are determined by the localized nonlinear physics rather than by the resolution of the full-order discretization, so the hyper-reduction acts as an adaptive integration strategy in space. The full-order mesh can be uniform and as fine as needed to resolve the physics, with no need to economize on elements or to adapt the mesh to the nonlinear regions beforehand, since the online cost barely grows with the refinement.

The remainder of the paper is organized as follows. Section~\ref{sec:model} presents the coupled hydro-mechanical model and Section~\ref{sec:fem} describes its finite element discretization. Section~\ref{sec:rom} develops the reduced-order methodology, including the field-split POD basis and the decomposed field-wise ECSW hyper-reduction. Section~\ref{sec:examples} presents the numerical examples, and Section~\ref{sec:concl} concludes the paper. 

\section{Coupled elastoplastic hydro-mechanical model}
\label{sec:model}

This section presents the equations that govern the coupled hydro-mechanical problem under isothermal conditions, following the classical formulations for deforming partially saturated porous media \cite{LewisSchrefler1998,ZienkiewiczCompGeo1999}.
\LL{To focus on the primary mechanisms governing the target problem, the following simplifying assumptions are adopted: (i) the gas permeability is infinitely large, maintaining gas pressure at constant atmospheric pressure, (ii) mass transfer between species is neglected, (iii) solid grains are considered incompressible, and (iv) quasi-static conditions apply (inertial effects are neglected).}
Two unknown fields are defined on the domain $\Omega\subset\mathbb{R}^d$, $d=2$ or $3$, and the time interval of interest: the solid displacement $\bu(\bx,t)$ and the pore water pressure $p(\bx,t)$. 
Throughout, a superposed dot denotes the time derivative, stresses are tension-positive, and the pore pressure is positive where the soil is saturated and negative (a suction) where it is unsaturated.

\subsection{\LL{Momentum balance equation and mechanical constitutive model}}
\label{sec:mech}

Neglecting inertia, balance of linear momentum reads
\begin{equation}
  \boldsymbol\nabla\!\cdot\!\bsig + \rho(p)\,\bg = \mathbf 0
  \qquad\text{in }\Omega, \label{eq:mom}
\end{equation}
where $\bsig$ is the total (Cauchy) stress, $\bg$ the gravity vector, and $\rho$ the bulk density. The density varies with the water content, and therefore with the pore pressure $p$, as
\begin{equation}
  \rho(p)=(1-n)\,\rho_s+\Theta(p)\,\rho_w, \label{eq:rho}
\end{equation}
where $n$ is the porosity, $\rho_s$ and $\rho_w$ the densities of the solid grains and of water, and $\Theta$ the volumetric water content.
The total stress is decomposed as
\begin{equation}
  \bsig=\bsig'-S_e(p)\,p\,\bI, \label{eq:bishop}
\end{equation}
with $\bsig'$ \LL{Bishop's stress tensor}, $S_e$ the effective degree of saturation, and $\bI$ the second-order identity tensor \cite{NuthLaloui2008}.
The water content $\Theta$ and the saturation $S_e$ are given by the retention model of Section~\ref{sec:flow}. Under the small-strain assumption, the strain tensor is the symmetric gradient of the displacement $\bu$,
\begin{equation}
  \beps(\bu)=\tfrac12\big(\boldsymbol\nabla\bu+\boldsymbol\nabla^{\!\top}\bu\big).
  \label{eq:strain}
\end{equation}
%Plane-strain conditions are assumed in the two-dimensional case.
%The effective stress $\bsig'$ is related to the strain through the elastoplastic law of Section~\ref{sec:dp}, while the water content $\Theta$ and the saturation $S_e$ are given by the retention model of Section~\ref{sec:flow}.

\LL{The stress-strain response is governed by an elastoplastic constitutive model, formulated in terms of Bishop's stress. }
The strain is split into elastic and plastic parts, $\beps=\beps^{e}+\beps^{p}$, \LL{ and stress changes are driven exclusively by elastic strains},
\begin{equation}
  \bsig'=\Cel:(\beps-\beps^{p}), \label{eq:hooke}
\end{equation}
where $\Cel$ is the fourth-order elastic stiffness tensor, taken isotropic and defined by Young's modulus $E$ and Poisson's ratio $\nu$.

The admissible stresses are bounded by a yield criterion $f\le0$, for which we adopt the Drucker--Prager model \cite{deSouzaNeto2008},
\begin{equation}
  f(\bsig')=\LL{\sqrt{3 J_2}}+3\,\eta\,p_\sigma-\sigma_Y,
  \label{eq:yield}
\end{equation}
\LL{where $p_\sigma=\tfrac13\tr\bsig'$ is Bishop's mean stress and $J_2=\tfrac12\,\mathbf s:\mathbf s$ the second invariant of the deviatoric stress $\mathbf s=\bsig'-p_\sigma\bI$.}
%where $p_\sigma=\tfrac13\tr\bsig'$ is Bishop's mean stress, and $J_2$ the second invariant of the deviatoric effective stress; $\eta$ and $\sigma_Y$ are the friction parameter and the yield strength. The plastic strain evolves through an associative flow rule,
\LL{Here, $\eta$ and $\sigma_Y$ are two constitutive parameters of the model.}
\LL{By matching the Mohr--Coulomb yield criterion with the inner (inscribed) Drucker--Prager cone \cite{deSouzaNeto2008}, these two parameters can be expressed in terms of the friction angle, $\phi$, and cohesion, $c$, as}
\begin{equation}
\eta=\frac{2\sin\phi}{3+\sin\phi},\qquad
\sigma_Y=\frac{6\,c\,\cos\phi}{3+\sin\phi}. \label{eq:dpmc}
\end{equation}
In the numerical examples of Section~\ref{sec:examples}, the
cohesion is fixed and the friction angle is one of the varying parameters, with $\eta$ and $\sigma_Y$ computed from \eqref{eq:dpmc} for each value of $\phi$.

\LL{Plastic strains evolve through an associated flow rule,}
\begin{equation}
  \dot\beps^{p}=\dot\lambda\,\frac{\partial f}{\partial\bsig'},
  \qquad \dot\lambda\ge0,\quad f\le0,\quad \dot\lambda\,f=0, \label{eq:flowrule}
\end{equation}
in which the plastic multiplier $\dot\lambda$ satisfies the loading--unloading (Karush--Kuhn--Tucker) conditions. We assume perfect plasticity, so that $\sigma_Y$ remains constant.

\LL{The} hyper-reduction strategy presented in Section~\ref{sec:rom} does not depend on this particular choice of constitutive model. The stress enters the formulation as a generic, history-dependent function of the strain, and the method never exploits the form of the yield surface or of the flow rule. Any other nonlinear constitutive law could be used without any modification to the reduction \LL{methodology}.
Drucker--Prager plasticity is adopted here only as a representative testbed.

\subsection{\LL{Mass balance equation and retention curve}}
\label{sec:flow}

%The pore air is assumed to remain at atmospheric pressure, so only the water flow is modelled. 
The mass balance of the pore water, combined with Darcy's law and the assumptions of incompressible solid grains and constant porosity in the storage terms, is expressed by the continuity equation
\begin{equation}
  \boldsymbol\nabla\!\cdot\!\Big(\tfrac{K(p)}{\gw}\big(\boldsymbol\nabla p-\rho_w\bg\big)\Big)
  -\Big(C(p)+\tfrac{\Theta(p)}{K_w}\Big)\dot p
  =\Theta(p)\,\boldsymbol\nabla\!\cdot\!\dot\bu
  \qquad\text{in }\Omega, \label{eq:flow}
\end{equation}
where $K$ is the hydraulic conductivity, $\gw=\rho_w g$ the unit weight of water, $C=\partial\Theta/\partial p$ the specific moisture capacity and $K_w$ the bulk modulus of water. 
The Darcy flux is defined as
\begin{equation}
  \q=-\tfrac{K(p)}{\gw}\big(\boldsymbol\nabla p-\rho_w\bg\big). \label{eq:darcy}
\end{equation}
The right-hand side of \eqref{eq:flow} couples the flow to the \LL{mechanical} problem, with the rate of volumetric strain $\boldsymbol\nabla\!\cdot\!\dot\bu$ acting as a source.

The \LL{soil-water} retention behaviour is described by the van Genuchten model \cite{vanGenuchten1980}, in which the effective saturation depends \LL{on the water pressure},
\begin{equation}
  S_e(p)=\left[1+\left(\alpha\left|\dfrac{p}{\gw}\right|\right)^{\frac{1}{1-m}}\right]^{-m}\ (p<0),
  \qquad S_e(p)=1\ (p\ge0), \label{eq:Se}
\end{equation}
with parameters $\alpha$ and $m$; the soil is saturated for $p\ge0$ and \LL{partially saturated for $p<0$}. 
\LL{The hydraulic conductivity depends on the effective degree of saturation through 
\begin{equation}
  K=K_s\,S_e^{3}, \label{eq:K}
\end{equation}
where $K_s$ is the saturated conductivity \cite{Charlier2013}. The water content is evaluated as
\begin{equation}
  \Theta=S_e(\Theta_s-\Theta_r)+\Theta_r, \label{eq:vG}
\end{equation}
where $\Theta_s$ is the saturated water content (equal to the porosity $n$) and $\Theta_r$ the residual water content.}
Through these relations the flow problem becomes strongly nonlinear in $p$.

\subsection{Initial and boundary conditions}\label{sec:bcs}

Each of the two problems, \LL{mechanical} and flow, has its own partition of the boundary $\partial\Omega$ into a Dirichlet, a Neumann and a Robin part. For the \LL{mechanical} problem,
\begin{equation}
  \bu=\hat\bu \ \text{ on }\Gamma_D^u,\qquad
  \bsig\!\cdot\!\bn=\hat{\mathbf t} \ \text{ on }\Gamma_N^u,\qquad
  \bsig\!\cdot\!\bn=-k_n\,(\bu\!\cdot\!\bn)\,\bn-k_t\,\bu_t
  \ \text{ on }\Gamma_R^u,
  \label{eq:bc-mech}
\end{equation}
with $\hat\bu$ a prescribed displacement, $\hat{\mathbf t}$ a prescribed traction and $\bn$ the outward unit normal. 
The last condition is a Winkler elastic foundation, in which the boundary reacts to the normal displacement $\bu\!\cdot\!\bn$ with modulus $k_n$ and to the tangential displacement $\bu_t=\bu-(\bu\!\cdot\!\bn)\,\bn$ with modulus $k_t$. 
This condition models the elastic support of the ground surrounding the computational domain. The structure can thus settle and slide against an elastic reaction instead of being rigidly clamped, which avoids the spurious stress concentrations that a hard support would induce.
For the flow problem,
\begin{equation}
  p=\hat p \ \text{ on }\Gamma_D^p,\qquad
  \q\!\cdot\!\bn=\hat q \ \text{ on }\Gamma_N^p,\qquad
  \q\!\cdot\!\bn=\beta\,\langle p\rangle \ \text{ on }\Gamma_R^p,
  \label{eq:bc-flow}
\end{equation}
with $\hat p$ a prescribed pressure and $\hat q$ a prescribed outflow. The last equation is a Robin condition that describes seepage, with the outflow proportional to the positive part of the pore pressure, $\langle p\rangle=\max(p,0)$, \LL{which allows outflow solely through saturated boundary segments.}
The conductance $\beta=K_s/(\gw\,\delta)$ depends on the saturated conductivity and on a characteristic seepage length $\delta$ \cite{Gerard2009}.
The specific boundary partitions and boundary data for the considered simulations are described in Section~\ref{sec:examples}.

The formulation is closed by the initial state $(\bu_0,p_0)$, obtained by solving the steady-state form of the coupled problem under the initial boundary data.

\section{Finite element full-order model}\label{sec:fem}

The coupled problem of Section~\ref{sec:model} is discretized in space with continuous $P_2$--$P_1$ finite elements and in time with the backward Euler scheme. This section states the choices made and introduces the notation of the resulting full-order model (FOM).
Special attention is paid to the nonlinearity of the residual and to its element-wise assembly, which are the two features exploited by the hyper-reduction of Section~\ref{sec:rom}.
Implementation of the model is done using the FEniCSx library \cite{fenicsx2023}.

\subsection{Weak form and time discretization}\label{sec:weak}

The momentum balance equation \eqref{eq:mom} and the mass balance equation \eqref{eq:flow} are written in weak form with test functions $\bv$ and $w$ for the displacement and the pressure, vanishing at the corresponding Dirichlet boundary, $
  \bv\in\mathcal V_0=\big\{\bv\in[H^1(\Omega)]^d \textnormal{ such that } \bv=\mathbf 0 \textnormal{ on }\Gamma_D^u\big\}$ and
  $
  w\in\mathcal W_0=\big\{w\in H^1(\Omega) \textnormal{ such that } w=0 \textnormal{ on }\Gamma_D^p\big\},$
with $d$ the space dimension.

Time is discretized with the backward Euler scheme. Over a step $[t^{n-1},t^{n}]$ of size $\Delta t$, the time derivatives are replaced by increments, $\dot\bu\approx(\bu^{n}-\bu^{n-1})/\Delta t$ and $\dot p\approx(p^{n}-p^{n-1})/\Delta t$, and the pressure-dependent coefficients $S_e$, $\Theta$, $C$, $K$ and $\rho$ are evaluated at the new pressure $p^{n}$, so that the problem is fully implicit. 

The problem to be solved at each time step is then to find $(\bu^{n},p^{n})\in[H^1(\Omega)]^d\times H^1(\Omega)$, satisfying the Dirichlet conditions in \eqref{eq:bc-mech} and \eqref{eq:bc-flow}, such that for all $(\bv,w)\in\mathcal V_0\times\mathcal W_0$,
\begin{align}
  r_u(\bu^{n},p^{n};\bv) &= \int_\Omega \bsig'\!:\!\beps(\bv)\,\mathrm{d}\Omega
   -\int_\Omega S_e\,p^{n}\,\boldsymbol\nabla\!\cdot\!\bv\,\mathrm{d}\Omega
   +\int_{\Gamma_R^u}\!\big(k_n\,(\bu^{n}\!\cdot\!\bn)(\bv\!\cdot\!\bn)
   +k_t\,\bu^{n}_t\!\cdot\!\bv_t\big)\,\mathrm{d}\Gamma
   \nonumber\\
  &\quad
   -\int_\Omega \rho\,\bg\!\cdot\!\bv\,\mathrm{d}\Omega
   -\int_{\Gamma_N^u}\hat{\mathbf t}\!\cdot\!\bv\,\mathrm{d}\Gamma = 0,
   \label{eq:wf-m}\\[0.3em]
  r_p(\bu^{n},p^{n};w) &= \int_\Omega \Theta\,\boldsymbol\nabla\!\cdot\!(\bu^{n}-\bu^{n-1})\,w\,\mathrm{d}\Omega
   +\int_\Omega\!\Big(C+\tfrac{\Theta}{K_w}\Big)(p^{n}-p^{n-1})\,w\,\mathrm{d}\Omega
   \nonumber\\
  &\quad
   +\Delta t\!\int_\Omega \tfrac{K}{\gw}\,\boldsymbol\nabla p^{n}\!\cdot\!\boldsymbol\nabla w\,\mathrm{d}\Omega
   +\Delta t\!\int_{\Gamma_R^p}\!\beta\,\langle p^{n}\rangle\,w\,\mathrm{d}\Gamma
   +\Delta t\!\int_{\Gamma_N^p}\!\hat q\,w\,\mathrm{d}\Gamma
   \nonumber\\
  &\quad
   -\Delta t\!\int_\Omega \tfrac{K}{\gw}\,\rho_w\bg\!\cdot\!\boldsymbol\nabla w\,\mathrm{d}\Omega = 0.
   \label{eq:wf-f}
\end{align}
Note that the problem is nonlinear on two accounts. \LL{Bishop's} stress $\bsig'$ is given by the elastoplastic law \eqref{eq:hooke}--\eqref{eq:flowrule}, a history-dependent function of the strain, and coefficients $\Theta$, $C$, $K$ in \eqref{eq:wf-f} depend on the unknown pressure $p^{n}$.

Both the \LL{water retention model} \eqref{eq:Se}--\eqref{eq:vG} and the positive part $\langle p\rangle$ of the seepage condition are defined piecewise, with a switch at $p=0$. In the implementation, both switches are regularized \LL{by a hyperbolic tangent with a characteristic width of one centimetre of pressure head}, which keeps the Jacobian bounded and continuous across the saturation front and on the seepage face.

\subsection{Spatial discretization}\label{sec:space}

The domain is partitioned into a mesh of triangles in two dimensions and of tetrahedra in three. 
The two unknown fields are approximated with inf-sup stable Taylor--Hood elements: quadratic for displacements and linear for pressure. 

Collecting the nodal unknowns at time $t^{n}$ in $\mathbf d^{n}=(\mathbf u^{n},\mathbf p^{n})\in\mathbb R^{N}$, with $n_u$ displacement and $n_p$ pressure degrees of freedom, $N=n_u+n_p$, the weak form \eqref{eq:wf-m}--\eqref{eq:wf-f} becomes a system of $N$ nonlinear algebraic equations,
\begin{equation}
  \mathbf r(\mathbf d^{n})=
  \begin{bmatrix}\mathbf r_u(\mathbf d^{n})\\[2pt]\mathbf r_p(\mathbf d^{n})\end{bmatrix}
  =\mathbf 0,
  \label{eq:fom}
\end{equation}
where the dependence on the previous converged state ($\mathbf d^{n-1}$ and the plastic variables stored at the quadrature points) is left implicit. 
As usual in the finite element method, the residual is assembled from element contributions,
\begin{equation}
  \mathbf r(\mathbf d)=\sum_{e\in\mathcal E}\mathbf L^{e\top}\,\mathbf r^e(\mathbf d),
  \label{eq:assembly}
\end{equation}
where $\mathcal E$ is the set of elements of the mesh, $\mathbf r^e$ the residual of element $e$, computed by integrating over the element, and $\mathbf L^e$ the Boolean matrix that gathers the degrees of freedom of $e$. The Jacobian is assembled in the same way, from the element Jacobians $\mathbf J^e$.
This sum-over-elements structure is precisely what the hyper-reduction of Section~\ref{sec:ecsw} will approximate by a short, weighted sum.

\subsection{Newton--Raphson solution of the time step}\label{sec:newton}

System \eqref{eq:fom} is solved by a monolithic Newton--Raphson method. 
Each iteration updates the solution as $\mathbf d^{n,i+1}=\mathbf d^{n,i}+(\delta\mathbf u,\delta\mathbf p)$, with increments given by
\begin{equation}
\begin{bmatrix} \mathbf J_{uu} & \mathbf J_{up}\\[2pt]
\mathbf J_{pu} & \mathbf J_{pp}\end{bmatrix}
\begin{bmatrix}\delta\mathbf u\\[2pt]\delta\mathbf p\end{bmatrix}
=-\begin{bmatrix}\mathbf r_u\\[2pt]\mathbf r_p\end{bmatrix},
\label{eq:newton}
\end{equation}
where $\mathbf J=\partial\mathbf r/\partial\mathbf d$ is the Jacobian of the residual, and all terms are evaluated at the current iterate $\mathbf d^{n,i}$.
Assembling the residual requires \LL{Bishop's stress} $\bsig'$, which follows from a local return-mapping problem, solved independently at each quadrature point and available in closed form for the Drucker--Prager model \cite{deSouzaNeto2008}.
Assembling the Jacobian requires the consistent tangent $\mathbb C_{\mathrm{tan}}=\partial\bsig'/\partial\beps$ of that same stress update, and the derivatives of the pressure-dependent coefficients with respect to $p^{n}$, obtained by differentiating the retention model \eqref{eq:Se}--\eqref{eq:vG}.
The expressions of the four blocks are collected in Appendix~\ref{app:jacobian}.
We iterate until the relative residual and the relative increments of both fields fall below $10^{-6}$.

%Displacements and pressures may differ by many orders of magnitude, so the mechanical unknowns and the mechanical residual are scaled by a constant factor before assembly, which balances the two blocks of \eqref{eq:newton} and keeps the monolithic system well conditioned. This is a change of units, and it is omitted from the notation.

The cost of the full-order model concentrates in two operations that are repeated at every Newton--Raphson iteration of every time step. The first is the assembly of the residual and the Jacobian, which loops over all the elements of the mesh and evaluates the constitutive law at every quadrature point. The second is the solution of the sparse linear system \eqref{eq:newton}. 
The reduced-order model of the next section is built to reduce both.

\section{Reduced-order model}\label{sec:rom}

The Reduced-Order Model (ROM) approximates the solution of the full-order model of Section~\ref{sec:fem} at a substantially lower computational cost. 
It is built offline from a training set of full-order simulations computed at sampled parameter values, with the solution at every time step stored as a snapshot. Online, for new parameter values, the solution is approximated by solving a reduced system with only a few unknowns, assembled over a small subset of the mesh.

We combine a field-split Proper Orthogonal Decomposition (POD) basis, which restricts the unknowns to a low-dimensional subspace (Section~\ref{sec:pod}), with the Energy-Conserving Sampling and Weighting method (ECSW), which defines a sparse cubature rule over selected elements (Section~\ref{sec:ecsw}).
In what follows, the model that uses the Galerkin projection onto the POD basis alone is referred to as the Reduced Basis (RB) model, and the model that, in addition, integrates the reduced equations with the ECSW rules as the Hyper-Reduced RB (HR-RB) model.
Both approximations are trained on the same snapshots, so no additional full-order simulations are needed, and their tolerances control the accuracy of the ROM. 
Two choices adapt the method to the particularities of the hydro-mechanical system: we train one independent ECSW rule per governing equation, and each rule approximates the nonlinear correction with respect to an affine baseline.
Section~\ref{sec:nnls} describes the algorithm in detail.

\subsection{Field-split POD reduced basis}\label{sec:pod}

The POD defines a low-dimensional linear subspace that approximates the manifold of solutions of the full-order model. 
The $N_s$ snapshots of the training set are collected in one matrix per field, $\mathbf X_u\in\mathbb R^{n_u\times N_s}$ and $\mathbf X_p\in\mathbb R^{n_p\times N_s}$, whose $N_s$ columns are the nodal values of the displacements and of the pore pressures of the training states, respectively. 
The snapshots are then centred by subtracting their mean, $(\bar{\mathbf u},\bar{\mathbf p})$, and each centred matrix is factorized by a Singular Value Decomposition (SVD). For the displacement, it reads
\begin{equation}
  \mathbf X_u-\bar{\mathbf u}\,\mathbf 1^{\!\top}
  =\mathbf U_u\,\boldsymbol\Sigma_u\,\mathbf V_u^{\!\top},
  \qquad
  \AM{\boldsymbol\Sigma_u=\operatorname{diag}(\AM{s_1,s_2,\dots}),
  \quad s_1\ge s_2\ge\cdots\ge0},
  \label{eq:svd}
\end{equation}
with $\mathbf 1\in\mathbb R^{N_s}$ the vector of ones. 
The columns of $\mathbf U_u$ form an orthonormal basis of the subspace spanned by the centred snapshots, ordered by decreasing singular value. In order to keep only the relevant information, the reduced basis $\mathbf B_u\in\mathbb R^{n_u\times k_u}$ is formed by the first $k_u$ columns of $\mathbf U_u$, retaining the modes that are significant relative to the first one,
\begin{equation}
  \AM{\frac{s_i}{s_1}}\ \ge\ \varepsilon_{\mathrm{POD}},
  \qquad i=1,\dots,k_u,
  \label{eq:podtrunc}
\end{equation}
for a prescribed truncation tolerance $\varepsilon_{\mathrm{POD}}$. 
The pressure basis $\mathbf B_p\in\mathbb R^{n_p\times k_p}$ is built analogously. 
Keeping the two fields separate avoids mixing different physical quantities that may differ in magnitude in a single SVD, so no cross-field normalization of the snapshots is needed. 

The reduced approximation is then
\begin{equation}
  \mathbf u \approx \bar{\mathbf u} + \mathbf B_u\,\mathbf z_u,\qquad
  \mathbf p \approx \bar{\mathbf p} + \mathbf B_p\,\mathbf z_p,
  \qquad\text{i.e.}\qquad
  \mathbf d \approx \bar{\mathbf d} + \mathbf B\mathbf z,
  \label{eq:rb}
\end{equation}
with $\bar{\mathbf d}=(\bar{\mathbf u},\bar{\mathbf p})$, $\mathbf B=\mathrm{diag}(\mathbf B_u,\mathbf B_p)$ block-diagonal and $\mathbf z=(\mathbf z_u,\mathbf z_p)\in\mathbb R^{k}$, $k=k_u+k_p\ll N$, the new unknowns.

The RB model is obtained by inserting \eqref{eq:rb} into the full-order residual \eqref{eq:fom} and testing with the basis. That is, we want to find $\mathbf z\in\mathbb R^{k}$ such that
\begin{equation}
  \mathbf r_r(\mathbf z)\,:=\,
  \mathbf B^{\!\top}\mathbf r\big(\bar{\mathbf d}+\mathbf B\mathbf z\big)=\mathbf 0.
  \label{eq:galerkin}
\end{equation}
This is a system of $k$ nonlinear equations, and it is solved by the same Newton--Raphson method as the full-order model. Each iteration solves the reduced counterpart of \eqref{eq:newton},
\begin{equation}
  \mathbf J_r\,\delta\mathbf z=-\,\mathbf r_r,
  \qquad\text{with}\quad
  \mathbf J_r=\mathbf B^{\!\top}\mathbf J\,\mathbf B,
  \label{eq:rom-newton}
\end{equation}
where the reduced Jacobian $\mathbf J_r\in\mathbb R^{k\times k}$ is a small dense matrix. 
\AM{In the examples of Section~\ref{sec:examples}, $k$ ranges from a few tens in two dimensions to about a hundred in three, while the full-order dimension $N$ goes from a few thousand to over half a million}.
The projection therefore makes the linear algebra inexpensive, but the assembly remains a bottleneck. Evaluating $\mathbf r_r$ and $\mathbf J_r$ still requires assembling $\mathbf r$ and $\mathbf J$ over the whole mesh, with the return mapping evaluated at every quadrature point. Moreover, the projection involves the 
products with $\mathbf B$ in \eqref{eq:galerkin}, whose cost also scales with the size of the mesh. 
The hyper-reduction of the next subsections further reduces these costs.

\subsection{Decomposed ECSW with an affine baseline}\label{sec:ecsw}

By the element-wise assembly \eqref{eq:assembly}, the reduced residual is a sum over the elements of the mesh $\mathcal E$,
\begin{equation}
  \mathbf r_r(\mathbf z)
  =\sum_{e\in\mathcal E}\mathbf B^{e\!\top}\,\mathbf r^e(\mathbf z),
  \qquad \mathbf B^e=\mathbf L^e\mathbf B.
  \label{eq:bsum}
\end{equation}
Here, $\mathbf B^e$ is the restriction of the reduced basis to element $e$, obtained with the Boolean matrix $\mathbf L^e$ that gathers its degrees of freedom.
Standard ECSW \cite{FarhatECSW2014} approximates this sum by a cubature over a small subset of elements $\tilde{\mathcal E}\subset\mathcal E$ as
\begin{equation}
  \mathbf r_r(\mathbf z)\;\approx\;
  \sum_{e\in\tilde{\mathcal E}}\xi^e\,\mathbf B^{e\!\top}\,\mathbf r^e(\mathbf z),
  \qquad \xi^e>0,\qquad
  |\tilde{\mathcal E}|\ll|\mathcal E|.
  \label{eq:ecsw}
\end{equation}
The reduced mesh $\tilde{\mathcal E}$ and the weights $\xi^e$ are determined offline, by imposing that the cubature reproduces the reduced residual on the training snapshots, as described below in Section~\ref{sec:nnls}. 
The same weights apply to the reduced Jacobian,
\begin{equation}
  \mathbf J_r(\mathbf z)\;\approx\;
  \sum_{e\in\tilde{\mathcal E}}\xi^e\,
  \mathbf B^{e\!\top}\,\mathbf J^e(\mathbf z)\,\mathbf B^e.
  \label{eq:ecsw-jac}
\end{equation}
The non-negativity of the weights preserves the structure of the Galerkin projection and makes the method robust, since symmetric positive semi-definite contributions remain so \cite{FarhatECSW2014,FarhatECSW2015}.

Two features of the present hydro-mechanical problem call for adapting this standard rule.
First, the nonlinearities of the two blocks act in different regions of the mesh. A single rule trained on the stacked residual would have to cover both regions with a common set of elements, each of them evaluated for both equations even where only one of the nonlinearities acts.
We therefore decompose the rule and train one cubature per governing equation, mirroring the field-split
structure of the POD basis and following
\cite{Zhang2025Decomposed,KehlsDecomposed2025}. The two rules
$(\tilde{\mathcal E}_u,\boldsymbol\xi_u)$ and
$(\tilde{\mathcal E}_p,\boldsymbol\xi_p)$ are trained independently, and the
two reduced meshes need not overlap.

Second, a significant part of the residual is not truly nonlinear.  
Each block residual is split into a baseline and a correction as
\[
  \mathbf r =\mathbf r_{\mathrm{base}}+\mathbf r_{\mathrm{corr}}.
\]
The baseline term contains terms that are affine in the state $(\bu,\mathbf p)$. It is computed exactly, and only the correction is approximated via ECSW.
In the baseline, the stress is the linear elastic response $\Cel\!:\!\beps(\bu)$, and the pressure-dependent coefficients ($K$, $S_e$, $\Theta$, $\rho$ and the storage coefficient) are evaluated at the mean pressure of the training snapshots, $\bar{\mathbf p}$, already available from the POD. 
Since these terms do not change with the solution, the exact projection of $\mathbf r_{\mathrm{base}}$ during the time loop reduces to a matrix-vector product of reduced dimension, between a matrix computed and stored before the time loop and the reduced-order solution, without assembly over the mesh.
%No linearity with respect to the parameters is required, since that assembly is performed once per parameter query and its cost is amortized over the time loop.
Keeping this part of the residual exact therefore preserves most of the computational gain, and confines the cubature error to the correction.
In our experience, the baseline-correction split makes the selected elements more localized; a numerical test quantifying this is reported in Section~\ref{sec:ex2d-results}.

With the two rules, and cubaturing only the correction, the residual blocks of the hyper-reduced model (HR-RB) read
\begin{equation}
\begin{aligned}
  \mathbf B_u^{\!\top}\mathbf r_u &\;\approx\;
  \underbrace{\mathbf B_u^{\!\top}\mathbf r_{u,\mathrm{base}}}_{\text{precomputed, exact}}
  \;+\;\sum_{e\in\tilde{\mathcal E}_u}\xi_u^e\,
  \mathbf B_u^{e\!\top}\,\mathbf r^e_{u,\mathrm{corr}},\\[6pt]
  \mathbf B_p^{\!\top}\mathbf r_p &\;\approx\;
  \underbrace{\mathbf B_p^{\!\top}\mathbf r_{p,\mathrm{base}}}_{\text{precomputed, exact}}
  \;+\;\sum_{e\in\tilde{\mathcal E}_p}\xi_p^e\,
  \mathbf B_p^{e\!\top}\,\mathbf r^e_{p,\mathrm{corr}},
\end{aligned}
  \label{eq:baseline}
\end{equation}
with $\mathbf B_u^e$, $\mathbf B_p^e$ the restrictions of the two bases to element $e$ and $\mathbf r^e_{u,\mathrm{corr}}$, $\mathbf r^e_{p,\mathrm{corr}}$ the element contributions to the corrections, the subscripts $u$ and $p$ denoting the block.
Each rule is thus trained to approximate the correction $\mathbf r_{\mathrm{corr}}$ of its own block.
The Jacobian is approximated analogously, with each rule weighting the row blocks of its own equation.

The constitutive update, and with it the plasticity return mapping, needs only to be solved at the quadrature points of the reduced meshes. The cost of evaluating the hyper-reduced residual and Jacobian therefore scales with $|\tilde{\mathcal E}_u|+|\tilde{\mathcal E}_p|$ and the reduced dimension $k$, and is independent of the size of the full mesh.

\AM{Note that, in the definition of the baselines, the nonlinear terms with respect to the displacement $\bu$ and the pressure $p$ are treated differently. The stress is taken as the linear elastic response, so the plastic contribution falls entirely to the correction, while the pressure-dependent coefficients are frozen at a reference state.
The stress depends on the strain history, so we cannot use a reference displacement as we do for the pressure.}
For the flow block, evaluating the coefficients at the mean pressure, rather than at full saturation, keeps the baseline accurate over the whole domain. A
saturated reference would be wrong by orders of magnitude in the hydraulic conductivity $K$ throughout the unsaturated zone, and the correction would then have to recover it.

Finally, only the volume integrals are hyper-reduced. The assembly of the boundary terms of
\eqref{eq:wf-m}-\eqref{eq:wf-f} (tractions, Winkler foundation and seepage) over boundary facets is inexpensive, so they are kept exact online.

\begin{remark}
In our FEniCSx implementation, the projection in the ECSW cubatures in \eqref{eq:baseline} is not performed at the element level.
The correction residual and Jacobian are assembled using the same variational forms as the full-order model, with two modifications that are native to the framework and that allow an easier implementation.
First, the integration measure is restricted to the reduced mesh $\tilde{\mathcal E}$, so that the assembly loops only over its elements, and a piecewise-constant function carrying the weights \AM{$\xi^e$} multiplies the integrands.
Second, the projection onto the reduced basis is performed afterwards, at once\AM{, through a truncated basis $\tilde{\mathbf B}$ that keeps only the rows associated with the degrees of freedom of $\tilde{\mathcal E}$.}
\end{remark}

\subsection{Element selection algorithm}\label{sec:nnls}

\AM{The cubature rules in \eqref{eq:baseline} are determined by the elements in the reduced meshes, $\tilde{\mathcal E}_u$ and $\tilde{\mathcal E}_p$, and the weights, $\xi_u^e$ and $\xi_p^e$. 
These are computed by imposing an accurate approximation of the correction residuals at the training snapshots.
Each of the two rules is obtained independently and in the same way, choosing the elements and the weights at the same time.
To simplify the notation, in this section we drop the block subscript ($u$ or $p$). }

%The cubature rules in \eqref{eq:baseline} are obtained by imposing an accurate approximation of the correction residuals at the training snapshots.
%Each of the two is computed independently and in the same way, choosing the elements of the reduced mesh and their weights at the same time. To simplify the notation, in this section we drop the block subscript ($u$ or $p$). 

\AM{
Let $\mathbf d_1,\dots,\mathbf d_{N_s}$ denote the snapshots in the training set.
For a given snapshot $\mathbf d_i$, the reduced correction can be expressed as a sum over all the elements of the mesh as
\begin{equation}
  \mathbf B^{\top}\mathbf r_{\mathrm{corr}}(\mathbf d_i)=\sum_{e\in\mathcal E}\mathbf B^{e\top}\mathbf r^e_{\mathrm{corr}}(\mathbf d_i).
  \label{eq:corrsum}
\end{equation}}
These element contributions are collected in a matrix
\begin{equation}
  \mathbf G=
  \begin{bmatrix}
    \mathbf B^{1\!\top}\mathbf r^{1}_{\mathrm{corr}}(\mathbf d_1) & \cdots &
    \mathbf B^{|\mathcal E|\!\top}\mathbf r^{|\mathcal E|}_{\mathrm{corr}}(\mathbf d_1)\\[4pt]
    \vdots & & \vdots\\[4pt]
    \mathbf B^{1\!\top}\mathbf r^{1}_{\mathrm{corr}}(\mathbf d_{N_s}) & \cdots &
    \mathbf B^{|\mathcal E|\!\top}\mathbf r^{|\mathcal E|}_{\mathrm{corr}}(\mathbf d_{N_s})
  \end{bmatrix},
  \label{eq:Gmatrix}
\end{equation}
where each entry is a vector with as many components as the reduced dimension $k$, so that $\mathbf G$ has $N_s\, k$ rows and $|\mathcal E|$ columns.
\AM{Adding the columns within the rows of snapshot $\mathbf d_i$ gives exactly $\mathbf B^{\top}\mathbf r_{\mathrm{corr}}(\mathbf d_i)$, so $\mathbf G\,\mathbf 1$ reproduces the corrections of the whole training set.}

The rule is then obtained by looking for non-negative weights $\boldsymbol\xi\in\mathbb R^{|\mathcal E|}$, with as few nonzero entries as possible, for which $\mathbf G\,\boldsymbol\xi$ approximates the target $\mathbf G\,\mathbf 1$ up to some tolerance $\tau$,
\begin{equation}
  \big\|\mathbf G\,\boldsymbol\xi-\mathbf G\,\mathbf 1\big\|
  \;\le\;\tau\,\big\|\mathbf G\,\mathbf 1\big\|,
  \qquad \boldsymbol\xi\ge\mathbf 0.
  \label{eq:nnls}
\end{equation}
The reduced mesh is then the support of the solution,
$\tilde{\mathcal E}=\{e \textnormal{ such that }\xi^e>0\}$.

Problem \eqref{eq:nnls} is solved by a greedy sparse Non-Negative
Least-Squares method (NNLS) of Lawson--Hanson type \cite{LawsonHanson1974,FarhatECSW2014}, summarized in Algorithm~\ref{alg:snnls}. Elements are added one at a time, in decreasing order of correlation with the current misfit, and the loop stops as soon as the criterion in \eqref{eq:nnls} is met. 
This early stop is what keeps the selection sparse,
since solving \eqref{eq:nnls} to optimality would return the trivial solution $\boldsymbol\xi=\mathbf 1$, the full mesh with unit weights. The tolerance $\tau$ therefore controls the trade-off between accuracy and sparsity.

The tolerance $\tau$ is independent of the POD truncation $\varepsilon_{\mathrm{POD}}$, since they control different errors. The POD tolerance sets the error of the reduced basis, and $\tau$ sets the integration error of the reduced operators. 
A larger tolerance $\tau$ gives a sparser rule but increases the cubature error. In practice, $\tau$ is chosen as large as possible while the HR-RB model remains on the error floor of the plain RB model.

\begin{algorithm}[htbp]
\caption{Greedy sparse NNLS selection of an ECSW rule, run separately for the
\LL{mechanical} and the flow block.}\label{alg:snnls}
\begin{algorithmic}[1]
\Require matrix $\mathbf G$, target $\mathbf G\,\mathbf 1$,
tolerance $\tau$
\State $\tilde{\mathcal E}\gets\emptyset$,\quad
$\boldsymbol\xi\gets\mathbf 0$,\quad $\mathbf m\gets\mathbf G\,\mathbf 1$ \Comment{rule and misfit initialization}
\While{$\|\mathbf m\|>\tau\,\|\mathbf G\,\mathbf 1\|$}
\State $e^{*}\gets\arg\max_{e\notin\tilde{\mathcal E}}
\big(\mathbf G^{\!\top}\mathbf m\big)_{e}$
\Comment{element most correlated with the misfit}
\State $\tilde{\mathcal E}\gets\tilde{\mathcal E}\cup\{e^{*}\}$
\State $\boldsymbol\zeta\gets\arg\min_{\boldsymbol\zeta\ge\mathbf 0}
\big\|\mathbf G_{\tilde{\mathcal E}}\,\boldsymbol\zeta-\mathbf G\,\mathbf 1\big\|$
\Comment{small NNLS on the selected columns $\mathbf G_{\tilde{\mathcal E}}$}
\State $\tilde{\mathcal E}\gets\{e\in\tilde{\mathcal E} \textnormal{ such that }\zeta^{e}>0\}$
\Comment{drop the elements zeroed by the solve}
\State $\boldsymbol\xi\gets\mathbf 0$;\quad
$\xi^{e}\gets\zeta^{e}\ \text{for }e\in\tilde{\mathcal E}$;\quad
$\mathbf m\gets\mathbf G\,(\mathbf 1-\boldsymbol\xi)$
\EndWhile
\State \Return $\boldsymbol\xi\ge\mathbf 0$ and
$\tilde{\mathcal E}=\{e \textnormal{ such that } \xi^{e}>0\}$
\end{algorithmic}
\end{algorithm}

\begin{remark}
Usually, in transient problems, matrix $\mathbf G \in \mathbb R^{N_s\, k \times |\mathcal E|}$ has many more rows than columns and is of large dimension.
The selection problem \eqref{eq:nnls}, although performed offline, can become expensive both in memory and in computational cost.
Here, both are alleviated by considering a QR factorization of the matrix, $\mathbf G=\mathbf Q\,\mathbf R$.
Since $\mathbf Q$ has orthonormal columns, replacing $\mathbf G$ by the triangular factor $\mathbf R \in \mathbb R^{|\mathcal E| \times |\mathcal E|}$, and the target by $\mathbf R\,\mathbf 1$, leaves every misfit norm unchanged. That is, we solve the equivalent problem
\begin{equation}
\big\|\mathbf R\,\boldsymbol\xi-\mathbf R\,\mathbf 1\big\|
\;\le\;\tau\,\big\|\mathbf R\,\mathbf 1\big\|,
\qquad \boldsymbol\xi\ge\mathbf 0,
\label{eq:nnls_QR}
\end{equation}
whose dimension is independent of the number of snapshots $N_s$.
In the implementation, the QR factorization is accumulated blockwise as the snapshots are processed, so that $\mathbf G$ is never stored.
This trick is sufficient for the meshes considered here, although $\mathbf R$ grows with the square of the number of elements. The size of the selection problem is a general concern, and other strategies have been proposed to keep it tractable, from compressing the training data beforehand, as in the empirical cubature method \cite{ECM2017}, to accelerated solvers for the non-negative least-squares problem \cite{Chapman2017,Grimberg2021}.
\end{remark}

\section{Numerical examples}\label{sec:examples}

The methodology is demonstrated on two numerical examples of earthfill dams retaining tailings and water. In both cases, we simulate the raise of the dam as a distributed load applied on the crest and upstream face of the domain, and the response is computed during the raising stage and the subsequent consolidation. 

The two setups are inspired by Nasika et al.~\cite{Nasika2022,Nasika2023}.
The first example considers a two-dimensional domain under plane-strain conditions, and is used to study the behaviour of the hyper-reduction under mesh refinement (Section~\ref{sec:ex2d}). The second example considers a dam in a valley, where the domain and the deposition process are fully three-dimensional, and is used to confirm that the conclusions carry over to a full-order model an order of magnitude larger (Section~\ref{sec:ex3d}).

The implementation uses FEniCSx/DOLFINx~0.9.0; the full-order systems are solved with the direct solver MUMPS~5.8 through PETSc~3.24, and Intel~MKL~2026.0 provides the BLAS and LAPACK kernels for both the factorizations and the dense operations of the reduced models.
Computational times are measured on a workstation equipped with two Intel Xeon Gold 6134 CPUs at 3.20\,GHz (8 cores per CPU) and 192\,GB of RAM, with all models running in a single process with 16 threads.

\subsection{Two-dimensional dam}\label{sec:ex2d}

\subsubsection{Problem setup}\label{sec:ex2d-setup}

\begin{figure}[htbp]
\centering
\includegraphics[width=.8\linewidth]{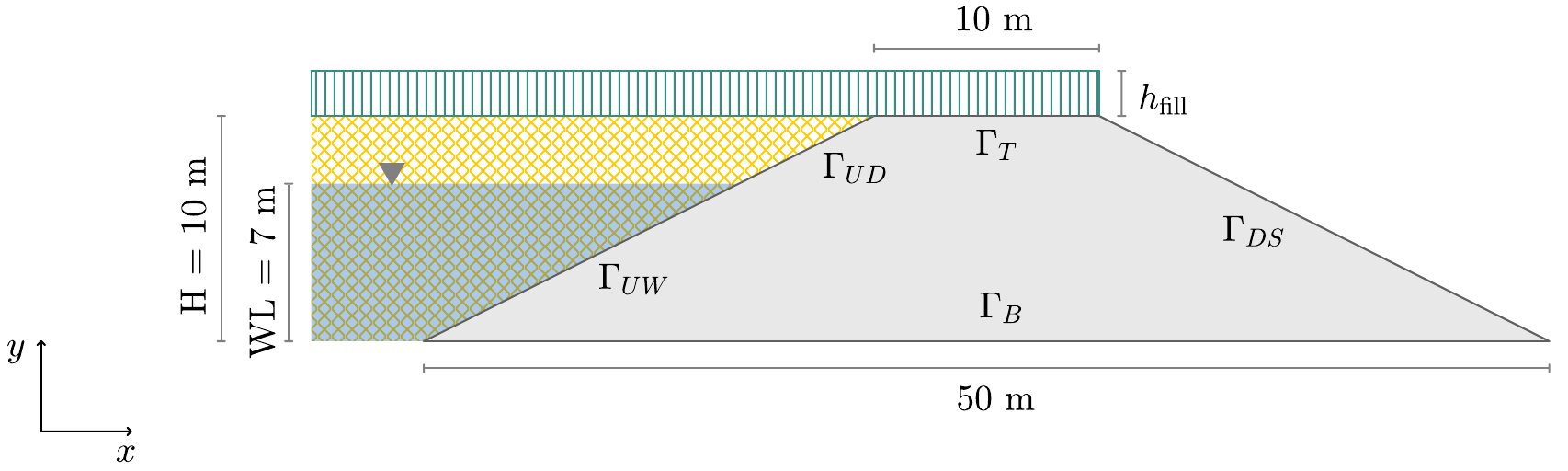}
\caption{Geometry of the two-dimensional dam, with its dimensions and the partition of the boundary into the wet and dry upstream faces $\Gamma_{UW}$ and $\Gamma_{UD}$, the crest $\Gamma_T$, the downstream face
$\Gamma_{DS}$ and the base $\Gamma_B$. The computational domain is the trapezoid in grey. Upstream, the retained tailings reach the crest elevation and the water the level $\mathrm{WL}=7\,\mathrm{m}$. The impoundment and the layer of fill of height $h_{\mathrm{fill}}$ that models the raise are not discretized, and act on the dam through the boundary conditions on $\Gamma_{UW}$, $\Gamma_{UD}$ and $\Gamma_T$.}
\label{fig:ex2d-domain}
\end{figure}

We consider an embankment of height $H=10\,\mathrm{m}$, with a crest of $10\,\mathrm{m}$ and a base of $50\,\mathrm{m}$. The computational domain is the symmetric trapezoid shown in Figure~\ref{fig:ex2d-domain}. 
The retained tailings
reach the crest elevation, with the water table at $\mathrm{WL}=7\,\mathrm{m}$. 
We simulate a raise of the dam of height $h_\mathrm{fill}$ and the subsequent consolidation. 
The impoundment is not discretized, and acts on the dam through the boundary conditions. The water level splits the upstream face into a wet part $\Gamma_{UW}$ ($y\le\mathrm{WL}$, with $y$ the elevation) and a dry part $\Gamma_{UD}$. The remaining boundaries are the crest $\Gamma_{T}$, the downstream face $\Gamma_{DS}$ and the base $\Gamma_{B}$.

\begin{table}[htbp]
	\centering
	\caption{Parameters of the dam, shared by the two examples. The first four
		are the parameters of the reduced-order model, sampled over the indicated
		ranges; the rest are fixed.\label{tab:ex2d-params}}
	\small
	\begin{tabular}{llll}
		\toprule
		symbol & description & value / range & sampling \\
		\midrule
		$\phi$ & friction angle of the dam body & $[28^{\circ},42^{\circ}]$ & linear \\
		$E$ & Young's modulus & $[10,250]\,\mathrm{MPa}$ & logarithmic \\
		$K_s$ & saturated conductivity & $[10^{-9},10^{-7}]\,\mathrm{m/s}$ & logarithmic \\
		$h_{\mathrm{fill}}$ & height of the raise & $[0.5,5]\,\mathrm{m}$ & linear \\
		\midrule
		$\nu$ & Poisson's ratio & $0.3$ & \\
		$c$ & cohesion & $15\,\mathrm{kPa}$ & \\
		$\eta$, $\sigma_Y$ & derived from $(\phi,c)$ by \eqref{eq:dpmc} & $0.27$--$0.36$, $22.9$--$18.2\,\mathrm{kPa}$ & \\
		$\phi_t$, $K_a$ & tailings friction angle, earth-pressure coeff. & $35^{\circ}$, $0.271$ & \\
		$\rho_s$, $\rho_w$ & solid and water densities & $2700$, $1000\,\mathrm{kg/m^3}$ & \\
		$K_w$ & water bulk modulus & $2.2\,\mathrm{GPa}$ & \\
		$\Theta_s\,(=n)$, $\Theta_r$ & saturated and residual water content & $0.38$, $0.038$ & \\
		$\alpha$, $m$ & van Genuchten parameters & $0.2\,\mathrm{m^{-1}}$, $0.184$ & \\
		$\gamma_t$, $\gamma_f$ & tailings and fill unit weights & $21$, $21\,\mathrm{kN/m^3}$ & \\
		$H$, $\mathrm{WL}$ & dam height and water level & $10$, $7\,\mathrm{m}$ & \\
		$k_n$, $k_t$ & foundation moduli (normal, tangential) & $10^{9}$, $10^{6}\,\mathrm{Pa/m}$ & \\
		$\delta$ & seepage length & $10\,\mathrm{m}$ & \\
		\bottomrule
	\end{tabular}
\end{table}

For the \LL{mechanical} problem, the dam is supported by the Winkler foundation of \eqref{eq:bc-mech}, with $\Gamma_R^u=\Gamma_B$, which models the compliance of the underlying ground. The normal modulus is large, $k_n=10^{9}\,\mathrm{Pa/m}$, so the ground acts as a nearly rigid support and the penetration of the dam into it is negligible. The tangential modulus is smaller, $k_t=10^{6}\,\mathrm{Pa/m}$, so the base can slide slightly over the ground.

The remaining faces carry prescribed tractions, $\Gamma_N^u=\Gamma_{UW}\cup\Gamma_{UD}\cup\Gamma_{T}\cup\Gamma_{DS}$. 
The upstream face is subjected to the loads of the tailings and of the water; the downstream face is traction-free.
The raise of the dam is modelled by a layer of fill of height $h_{\mathrm{fill}}$ and unit weight $\gamma_f$ placed on top of the domain, and acts as a load on the crest and, through the tailings, on the upstream face. The raise is ramped linearly over $T_r=10$ days,
\begin{equation}
  q_y(t)=\psi(t)\,\gamma_f\,h_{\mathrm{fill}},\qquad
  q_x=K_a\,q_y,\qquad
  \psi(t)=\min\!\big(t/T_r,\,1\big),
  \label{eq:ex2d-ramp}
\end{equation}
where $K_a=(1-\sin\phi_t)/(1+\sin\phi_t)$ is the active earth-pressure coefficient of the retained tailings, computed from their friction angle $\phi_t=35^{\circ}$. 
Note that the tailings and the dam body are different materials, and the friction angle of the dam body is denoted by $\phi$. 
The tractions read
\begin{align}
  \hat{\mathbf t}&=\big(q_x+K_a\,\sigma_v',\;-q_y-\sigma_v'\big)^{\!\top}
    &&\text{on }\Gamma_{UD},\nonumber\\
  \hat{\mathbf t}&=\big(q_x+K_a\,\sigma_v',\;-q_y-\sigma_v'\big)^{\!\top}
    -p_w\,\bn
    &&\text{on }\Gamma_{UW},\label{eq:ex2d-tractions}\\
  \hat{\mathbf t}&=\big(0,\;-q_y\big)^{\!\top}
    &&\text{on }\Gamma_{T},\nonumber \\
    \hat{\mathbf t}&=\big(0,\;0\big)^{\!\top}
    &&\text{on }\Gamma_{DS},\nonumber
\end{align}
with $\bn$ the outward unit normal, $p_w=\gw(\mathrm{WL}-y)$ the water pressure and $\sigma_v'$ the vertical effective stress in the tailings, that is, the total weight of the tailings above the water level plus their submerged weight below it,
\begin{equation}
  \sigma_v'(y)=
  \begin{cases}
    \gamma_t\,(H-y), & y>\mathrm{WL},\\[2pt]
    \gamma_t\,(H-\mathrm{WL})+(\gamma_t-\gw)\,(\mathrm{WL}-y), &
    y\le\mathrm{WL},
  \end{cases}
  \label{eq:ex2d-sigmav}
\end{equation}
where $\gamma_t$ is the unit weight of the tailings. %All load magnitudes are defined positive; with the tension-positive convention, the signs in \eqref{eq:ex2d-tractions} make every traction compressive.

For the flow problem, the wet upstream face is a Dirichlet boundary, $\Gamma_D^p=\Gamma_{UW}$, with prescribed hydrostatic pressure $\hat p=\gw(\mathrm{WL}-y)$. 
The dry upstream face, the crest and the base are impervious, with $\hat q=0$ on $\Gamma_N^p=\Gamma_{UD}\cup\Gamma_{T}\cup\Gamma_{B}$. The downstream face is a seepage face as in \eqref{eq:bc-flow}, $\Gamma_R^p=\Gamma_{DS}$, with seepage length $\delta=10$ m.

All material and loading parameters are listed in Table~\ref{tab:ex2d-params}. Four of them are varied in the ROM, and the parametrization is discussed in Section~\ref{sec:ex2d-train}.

The initial condition is the steady state of the dam before the raise, corresponding to $q_y=0$. 
From this state, the transient simulation covers 30 days, comprising 10 days of raising followed
by 20 days of consolidation \LL{phase}. Time is discretized with 312 backward Euler steps. The time step is
$\Delta t=0.05$ days up to day 12, which resolves the raise and the early
consolidation, and $\Delta t=0.25$ days thereafter.
In space, the dam is discretized using three unstructured triangular meshes of increasing refinement, all with $P_2$--$P_1$ Taylor--Hood elements (Table~\ref{tab:ex2d-meshes}). Element integrals are computed with a quadrature rule of order five, with seven points per triangle.

\begin{table}[htbp]
\centering
\caption{Number of elements $|\mathcal E|$, quadrature points, displacement
and pressure degrees of freedom, and total number of unknowns $N$, for the
three meshes of the two-dimensional dam.\label{tab:ex2d-meshes}}
\begin{tabular}{lccccc}
\toprule
mesh & $|\mathcal E|$ & quad.\ points & $n_u$ & $n_p$ & $N = n_u + n_p$ \\
\midrule
coarsest     & 990    & 6\,930  & 4\,206  & 557    & 4\,763 \\
intermediate & 2\,963 & 20\,741 & 12\,276 & 1\,588 & 13\,864 \\
finest       & 9\,982 & 69\,874 & 40\,718 & 5\,189 & 45\,907 \\
\bottomrule
\end{tabular}
\end{table}

\subsubsection{Parameters and training set}\label{sec:ex2d-train}

The parameters of the ROM are the friction angle of the dam body $\phi$, its Young's modulus $E$, the saturated conductivity $K_s$ and the height of the raise $h_{\mathrm{fill}}$. The first three control the strength of the dam, its deformability and the dissipation rate of the pore pressure, and their value is subject to significant uncertainty in practice. The strength and the stiffness of the fill depend on its compaction and deposition history, and are rarely known with precision; the conductivity is highly sensitive to compaction, and can vary by orders of magnitude for the same material \cite{Whalley2012,Nasika2023}. The height of the raise is not a material property, but an operational parameter that sets the magnitude of the load and varies from one construction scenario to another.

The parameters are sampled over the ranges in Table~\ref{tab:ex2d-params}.
The range $\phi\in[28^{\circ},42^{\circ}]$ spans fill states from loose and slimes-like to dense and compacted. Similarly, the modulus $E\in[10,250]\,\mathrm{MPa}$ goes from soft, lightly compacted fills to stiff, well-compacted ones, and the conductivity $K_s\in[10^{-9},10^{-7}]\,\mathrm{m/s}$ from almost impervious fills, which consolidate over months, to more pervious ones, which drain within
days. 
The selected ranges are consistent with values reported for earthfill materials \cite{Pinyol2008,Nasika2023}. 
Parameters $E$ and $K_s$ are sampled logarithmically, as they span more than one and two orders of magnitude. 
The ranges are wide on purpose, so that the responses differ strongly; the ROM must capture this variability to be accurate over the whole parameter range.

In the training set, we include the simulations from $40$ points in the parameter space, selected by Latin Hypercube Sampling. Each training simulation contributes its 313 states (the initial condition and the 312 time steps) as snapshot columns, giving $N_s=12\,520$ snapshots per field. Ten additional points are drawn to be used as an independent test set to validate the ROM.

\begin{figure}[htbp]
\centering
\includegraphics[width=\linewidth]{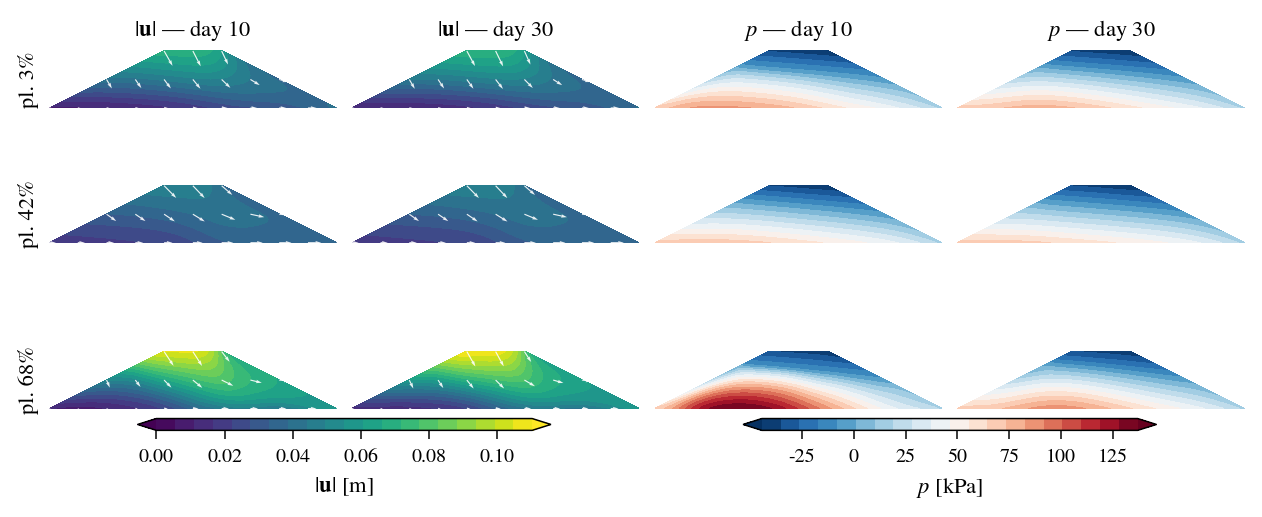}
\caption{Three training simulations on the finest mesh, illustrating the variability of the training set. Each row corresponds to one parameter point $(\phi, E, K_s, h_{\mathrm{fill}})$; from top to bottom, $(37.1^{\circ},\,13\ \mathrm{MPa},\,10^{-9}\ \mathrm{m/s},\,1.0\ \mathrm{m})$, $(31.8^{\circ},\,40,\,4.6\times10^{-8},\,3.7)$ and
$(28.1^{\circ},\,15,\,2.4\times10^{-9},\,4.6)$, with $3\%$, $42\%$ and $68\%$ of the quadrature points plastified at the end of the simulation.
The panels show the displacement magnitude (left pair of columns) and the pore pressure (right pair), at the end of the raise (day 10) and at the end of the simulation (day 30).}\label{fig:ex2d-sols}
\end{figure}

\begin{figure}[htbp]
\centering
\includegraphics[width=\linewidth]{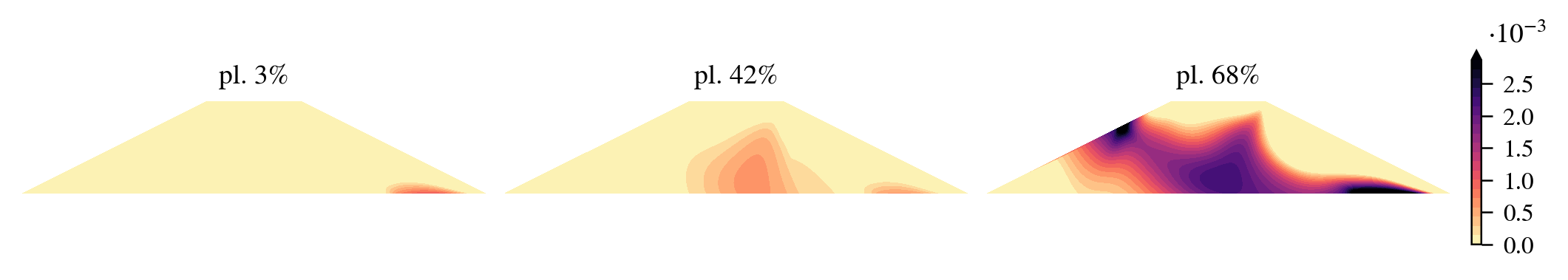}
\caption{\LL{Accumulated plastic multiplier} at the end of the simulation (day 30) for the three parameter points of Figure~\ref{fig:ex2d-sols}, with $3\%$, $42\%$ and $68\%$ of the quadrature points plastified.}\label{fig:ex2d-alpha}
\end{figure}

Figures~\ref{fig:ex2d-sols} and~\ref{fig:ex2d-alpha} illustrate the variability in the family of solutions of the problem. In the training set, at the end of the simulation, the fraction of quadrature points that \LL{have been subjected to plastic strains} ranges from $3\%$ to $77\%$. 
%The most plastified responses combine a low friction angle $\phi$ with a high raise $h_{\mathrm{fill}}$. 
\LL{Naturally, the responses with the largest regions of plastic yielding correspond to cases combining a low friction angle $\phi$ with a large fill height raise $h_{\mathrm{fill}}$.}

\subsubsection{Reduced bases and reduced meshes}\label{sec:ex2d-rom}

The POD bases are computed from the training sets for the displacements and the pressure, with a tolerance $\varepsilon_{\mathrm{POD}}=10^{-3}$ in the criterion \eqref{eq:podtrunc} for both fields. This gives $k_u=13$ displacement modes and $k_p=21$ pressure modes on the three meshes. 
The reduced problem has thus 34 unknowns, in place of the $4\,763$ to $45\,907$ of the full-order models. 
Note that the reduced dimension does not change with the mesh, since the number of modes is set by the variability in the family of solutions and its parametrization rather than by the spatial discretization, provided every mesh resolves the physics. 
The gain is therefore more relevant for the finest mesh, as the percentages in Table~\ref{tab:ex2d-rom} show.

\begin{table}[htbp]
\centering
\caption{Number of modes in the POD bases (with truncation tolerance $\varepsilon_{\mathrm{POD}}=10^{-3}$) and
decomposed ECSW rules (with tolerance $\tau=10^{-2}$) on the three meshes. For the bases, the percentages are relative to the number of degrees of freedom of each field, $n_u$ and $n_p$; for the rules, to the total number of elements $|\mathcal E|$.\label{tab:ex2d-rom}}
\begin{tabular}{ccccc}
\toprule
$|\mathcal E|$ & $k_u$ & $k_p$ & $|\tilde{\mathcal E}_u|$ & $|\tilde{\mathcal E}_p|$ \\
\midrule
990    & 13\ \,(0.31\%) & 21\ \,(3.8\%) & 48\ \,(4.8\%)  & 64\ \,(6.5\%) \\
2\,963 & 13\ \,(0.11\%) & 21\ \,(1.3\%) & 57\ \,(1.9\%)  & 68\ \,(2.3\%) \\
9\,982 & 13\ \,(0.03\%) & 21\ \,(0.4\%) & 55\ \,(0.6\%)  & 75\ \,(0.8\%) \\
\bottomrule
\end{tabular}
\end{table}

The decomposed ECSW rules are trained with the same tolerance $\tau=10^{-2}$ in the criterion \eqref{eq:nnls} for the two blocks, and Table~\ref{tab:ex2d-rom} reports the number of selected elements in the reduced meshes.
This tolerance is sufficient, in the sense that the errors of the HR-RB model match those of the RB model, so the integration approximation does not degrade the accuracy of the reduction, as shown later in
Section~\ref{sec:ex2d-results}.

\begin{figure}[htbp]
\centering
\includegraphics[width=\linewidth]{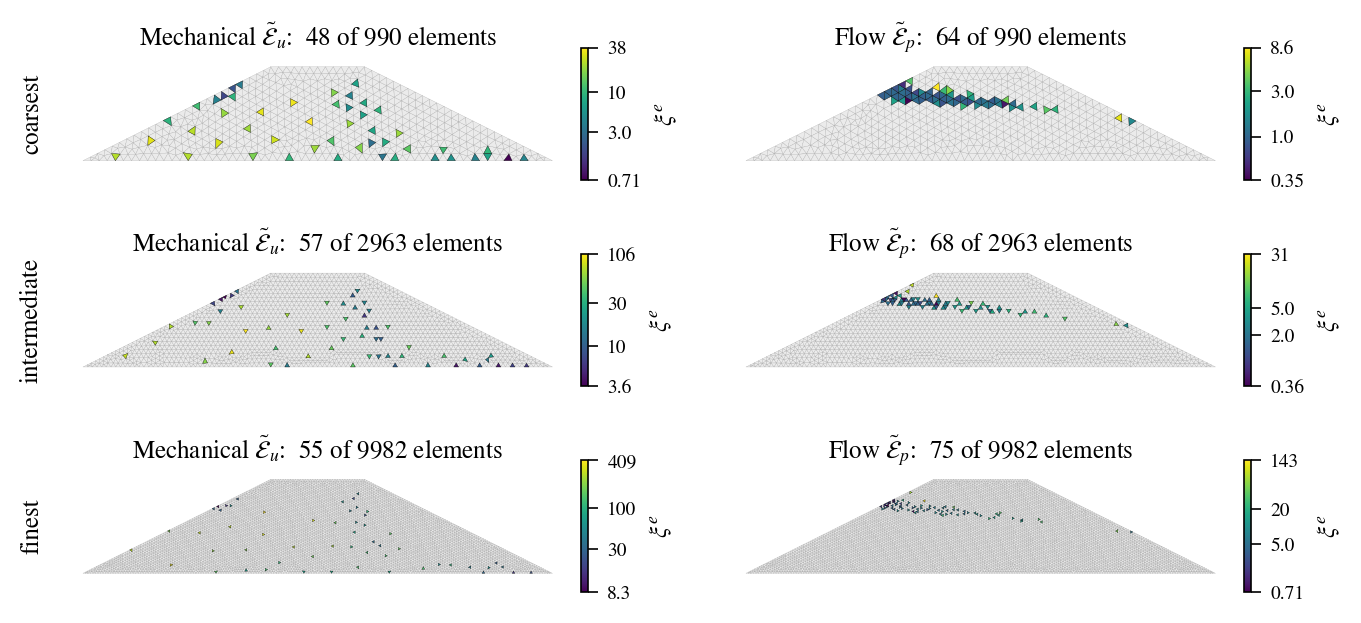}
\caption{Decomposed ECSW rules on the three meshes: \LL{mechanical} rules $\tilde{\mathcal E}_u$ (left) and flow rules $\tilde{\mathcal E}_p$ (right), coloured by their weights $\xi^e$ on a logarithmic colour scale, with the full mesh in grey. The \LL{selected elements in $\tilde{\mathcal E}_u$} spread over the region swept by the plastic zones of the training set, and the \LL{elements in $\tilde{\mathcal E}_p$} concentrate along the phreatic \LL{surface}. The rules occupy the same regions on every mesh, with numbers of elements that grow far more slowly than the mesh.}\label{fig:ex2d-rules}
\end{figure}

Figure~\ref{fig:ex2d-rules} shows where the rules place their elements on the three meshes, and that the two reduced meshes do not overlap. The elements are placed where the nonlinearity acts. 
The \LL{elements selected in $\tilde{\mathcal E}_u$} spread over the region swept by the plastic zones across the training set, with a concentration towards the downstream toe and the base, whereas \LL{those in $\tilde{\mathcal E}_p$} concentrate \LL{in a narrow band along the phreatic surface}. The same regions are selected on every mesh, and the reduced meshes barely grow under refinement. While the element count grows tenfold, from $990$ to $9\,982$, $|\tilde{\mathcal E}_u|$ only goes from 48 to 55 and $|\tilde{\mathcal E}_p|$ from 64 to 75, both
essentially saturating. As fractions of the mesh, the rules shrink from $4.8\%$ and $6.5\%$ to $0.6\%$ and $0.8\%$. 
\AM{
The selection algorithm tracks the regions where the soil yields and desaturates, and the extent of these regions is set by the physics rather than by the resolution.
}

\paragraph*{\AM{Effect of the affine baseline: concentration of the selection.}}
The presented ECSW rules are trained only on the correction $\mathbf r_{\mathrm{corr}}$ of the residual with respect to an affine, precomputed baseline, following the split in \eqref{eq:baseline}.
To analyse the effect of the split in the rules, we trained a control rule for the coarsest mesh directly on the full residual $\mathbf r$.
In this case, matrix $\mathbf G$ is computed using $\mathbf r$ itself at the training snapshots. The stopping tolerance $\tau$ in criterion \eqref{eq:nnls} is relative to the target of the approximation, so it has to be rescaled for the two rules to be comparable. 
For the \LL{mechanical} block, $\|\mathbf B_u^{\!\top}\mathbf r_{u,\mathrm{corr}}\| \approx 0.058\,\|\mathbf B_u^{\!\top}\mathbf r_u\|$ over the training snapshots, and the tolerance of the control is set to $\tau'_u=0.058\,\tau=5.80\cdot10^{-4}$. Analogously, the ratio for the flow block is $0.453$, giving $\tau'_p=4.53\cdot10^{-3}$. The two rules then target the same absolute accuracy.

\begin{figure}[htbp]
\centering
\includegraphics[width=\linewidth]{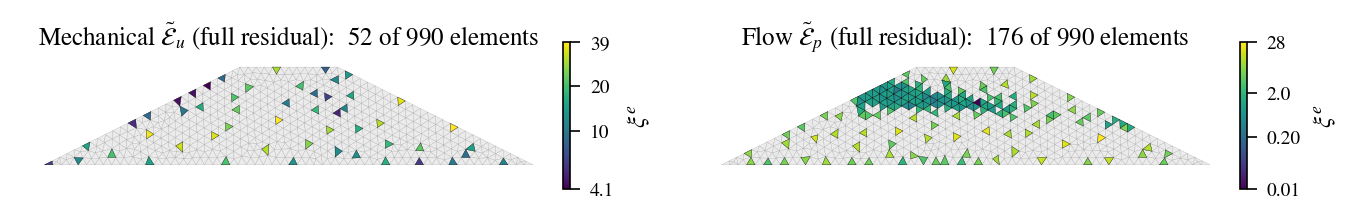}
\caption{ECSW rules trained on the full residual on the
coarsest mesh, at the matched tolerances $\tau'_u$ and $\tau'_p$, coloured
by their weights $\xi^e$ on a logarithmic colour scale. Without
the baseline, both selections spread over the whole dam: the rules must
also represent the affine part of the residual, which the baseline
otherwise computes exactly and at no cost. Compare with the first row of
Figure~\ref{fig:ex2d-rules}.}\label{fig:ex2d-fullres}
\end{figure}

At this matched tolerance, the effect of splitting the residual differs between the two blocks. Table~\ref{tab:ablation} compares the two options, with and without the affine baseline, indicating the number of selected elements and the error in the approximation of the reduced residual on the test simulations. The selected elements are plotted in
Figure~\ref{fig:ex2d-fullres}.

The flow reduced mesh grows from $64$ to $176$ elements for an error of the same order. The selection no longer confines itself \LL{near the phreatic surface} and spreads over the whole dam, where no nonlinearity acts. The reason is that, without the split, the rule itself
must represent the mean-state response, which the affine baseline computes exactly.

In contrast, the \LL{mechanical} reduced mesh grows only marginally, from $48$ to $52$ elements, with comparable accuracy. The baseline brings little gain in size for this block for two reasons. First, the elastic response is itself cheap to represent with a sparse cubature. Second, there is no small region for the selection to confine to, since the plastic zone
varies strongly across the training set, spanning $3\%$ to $77\%$ of the quadrature points depending on the sample, and both rules must cover a large part of the dam.

Overall, training on the correction drives the selection to the regions where the nonlinearity acts, and the gain is therefore largest when these regions are localized. The baseline-corrected rules are the ones used in the remainder of the paper.

\begin{table}[htbp]
\centering
\caption{Effect of the affine baseline, on the coarsest mesh: number of
selected elements and relative error in the approximation of the reduced
residual on the ten test simulations (mean / max over their time steps).
The error is measured in the Euclidean norm and is relative to the norm
of the reduced residual for both rules, so they are directly comparable. The
baseline-corrected rules use the tolerance $\tau=10^{-2}$, and the
full-residual ones the matched tolerances $\tau'_u$ and
$\tau'_p$.\label{tab:ablation}}
\begin{tabular}{llcc}
\toprule
block & rule & $|\tilde{\mathcal E}|$ & residual error (mean / max) \\
\midrule
mechanical & baseline-corrected ($\tau$)      & 48  & $4.7\times10^{-4}$ / $1.1\times10^{-3}$ \\
     & full residual ($\tau'_u$)        & 52  & $5.2\times10^{-4}$ / $1.1\times10^{-3}$ \\
\midrule
flow & baseline-corrected ($\tau$)      & 64  & $1.6\times10^{-2}$ / $1.0\times10^{-1}$ \\
     & full residual ($\tau'_p$)        & 176 & $2.3\times10^{-2}$ / $1.7\times10^{-1}$ \\
\bottomrule
\end{tabular}
\end{table}

\subsubsection{Accuracy and speed-up}\label{sec:ex2d-results}

The accuracy of the ROM simulations is measured with two errors with respect to the full-order solutions. The first one is a global error over the whole simulation,
\begin{equation}
e_u=\Bigg(
\frac{\sum_n\big\|\mathbf u^{n}_{\mathrm{ROM}}-\mathbf u^{n}_{\mathrm{FOM}}\big\|^{2}}
{\sum_n\big\|\mathbf u^{n}_{\mathrm{FOM}}\big\|^{2}}
\Bigg)^{\!1/2},
\qquad\text{and analogously } e_p,
\label{eq:ex2d-err}
\end{equation}
where $n$ runs over the time steps of the simulation. The second one is the maximum error over the time steps,
$e_u^{\mathrm{max}}=\max_n
\|\mathbf u^{n}_{\mathrm{ROM}}-\mathbf u^{n}_{\mathrm{FOM}}\|/
\|\mathbf u^{n}_{\mathrm{FOM}}\|$.

\begin{figure}[htbp]
\centering
\includegraphics[width=.95\linewidth]{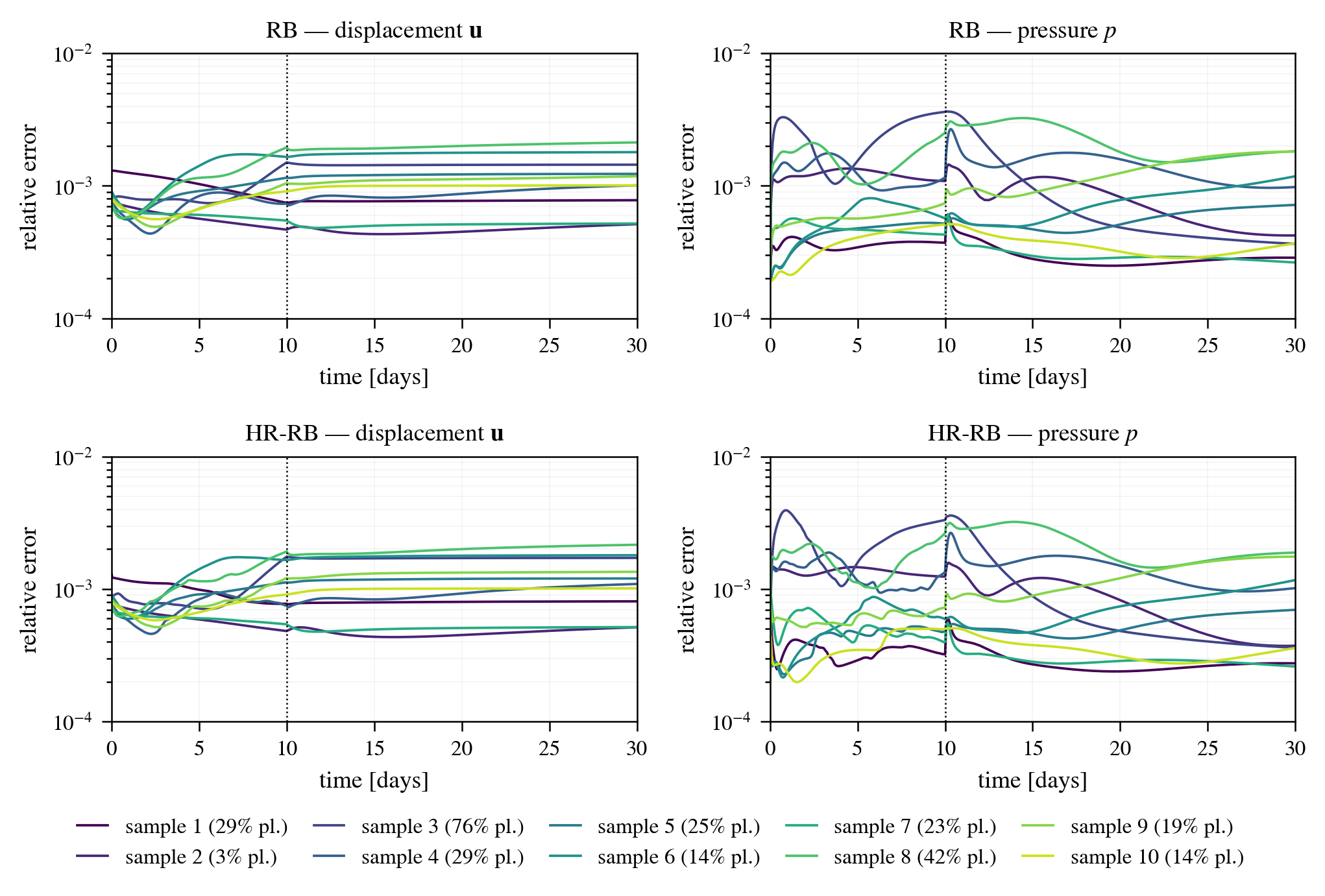}
\caption{Evolution in time of the per-step relative errors on the ten test samples of the finest mesh, for the RB (top row) and the HR-RB (bottom row) models, with the displacement on the left and the pressure on the right. The legend indicates the fraction of quadrature points plastified at the end of each simulation, and the dotted line
marks the end of the raise at day 10. Samples 1 to 10 correspond to
$(\phi, E, K_s, h_{\mathrm{fill}})=
(28.7^{\circ},\,29\ \mathrm{MPa},\,1.4\times10^{-8}\ \mathrm{m/s},\,1.2\ \mathrm{m})$,
$(35.4^{\circ},\,16,\,1.8\times10^{-9},\,0.5)$,
$(30.9^{\circ},\,136,\,2.0\times10^{-9},\,3.4)$,
$(34.0^{\circ},\,17,\,9.5\times10^{-9},\,3.5)$,
$(38.5^{\circ},\,43,\,5.0\times10^{-8},\,4.4)$,
$(41.0^{\circ},\,33,\,6.6\times10^{-8},\,4.5)$,
$(40.5^{\circ},\,81,\,5.5\times10^{-9},\,2.2)$,
$(31.5^{\circ},\,13,\,1.5\times10^{-8},\,4.1)$,
$(32.1^{\circ},\,12,\,4.2\times10^{-8},\,2.6)$ and
$(36.8^{\circ},\,31,\,3.2\times10^{-8},\,1.9)$.
The errors on the coarsest and intermediate meshes are very similar to these ones. \AM{The errors stay near $10^{-3}$ throughout the raise and the consolidation, and the HR-RB curves are indistinguishable from the RB ones.}}\label{fig:ex2d-err}
\end{figure}

Figure~\ref{fig:ex2d-err} shows the evolution in time of the per-step relative errors of the two reduced models, RB and HR-RB, on the ten test simulations of the finest mesh, and Table~\ref{tab:ex2d-errors} collects the mean errors on the three meshes.
Both reduced models reproduce both fields to about $0.1\%$ in the global error, with maximum errors below $0.2\%$, and the errors stay around $10^{-3}$ through the raise
and the consolidation. 
The RB and HR-RB results are essentially identical, so the ECSW cubature adds no visible error on top of the basis truncation with the chosen tolerances. 
The error levels are also very similar on the three meshes, since they are controlled by  $\varepsilon_{\mathrm{POD}}$ rather than by the mesh size.

\begin{table}[htbp]
\centering
\caption{Mean relative errors over the 10 test simulations --- global errors \eqref{eq:ex2d-err} and maximum errors --- for the RB and HR-RB models on the three meshes. \AM{Reduced models reproduce both fields to about $0.1\%$ in the global error. The RB and HR-RB results are essentially identical, so ECSW adds no error on top of the basis truncation.}\label{tab:ex2d-errors}}
\begin{tabular}{clcccc}
\toprule
$|\mathcal E|$ & model & $e_u$ & $e_p$ & $e_u^{\mathrm{max}}$ & $e_p^{\mathrm{max}}$ \\
\midrule
990    & RB    & 1.03e-3 & 9.73e-4 & 1.28e-3 & 1.61e-3 \\
       & HR-RB & 1.07e-3 & 1.00e-3 & 1.34e-3 & 1.72e-3 \\
\midrule
2\,963 & RB    & 1.03e-3 & 1.00e-3 & 1.28e-3 & 1.65e-3 \\
       & HR-RB & 1.07e-3 & 1.04e-3 & 1.34e-3 & 1.80e-3 \\
\midrule
9\,982 & RB    & 1.01e-3 & 1.01e-3 & 1.26e-3 & 1.65e-3 \\
       & HR-RB & 1.04e-3 & 1.03e-3 & 1.31e-3 & 1.71e-3 \\
\bottomrule
\end{tabular}
\end{table}

Table~\ref{tab:ex2d-timing} reports wall-clock times for the same three test
simulations on each mesh, chosen to span the plastification range ($3\%$,
$42\%$ and $76\%$ of the quadrature points plastified).
The RB model alone achieves modest factors, between $1.0\times$ and $1.4\times$, that do not improve under refinement. The reduced basis decreases the solve time, but the assembly and the return mapping still run over the full mesh. Moreover, as anticipated in Section~\ref{sec:pod}, the solve time of the RB model is dominated not by the dense $k\times k$ system, but by the projection of the assembled residual and Jacobian onto the basis, whose cost also scales with the
full mesh.

The ECSW hyper-reduction strategy significantly reduces this cost. The return mapping is evaluated only at the quadrature points of $\tilde{\mathcal E}_u$, the assembly loops only
over the reduced meshes, and the projection involves only the restriction of the modes to them. 
The online stage of the HR-RB model includes the precomputation of the reduced baseline operators, which takes between $0.3$ and $2.4\,\mathrm{s}$ and is counted in the totals of Table~\ref{tab:ex2d-timing}. The resulting speed-ups are $7$--$9\times$ on
the coarsest mesh, $14$--$19\times$ on the intermediate one and $30$--$43\times$ on the finest one.
The gain thus increases with refinement. Refinement makes the full-order model more expensive, while the hyper-reduced online cost, which is set by the sizes of the bases and of the reduced meshes, grows far more slowly. In practice, this means there is no need to be sparing with the full-order discretization. The mesh can be uniform and as fine as needed to resolve the physics, and the ECSW selection acts as an adaptive strategy that concentrates the online effort on the plastified zones and the phreatic \LL{surface}.

\LL{The speed-up also grows with the extent of the region undergoing plastic deformation. A response with larger plastic zones demands more Newton--Raphson iterations per step and a larger number of evaluations of the plastic return mapping, and this is precisely the nonlinear computational work that the hyper-reduction bypasses.}
In all cases, the number of Newton--Raphson iterations of the reduced models stays within $16\%$ of the full-order count, so the reduction does not degrade the convergence of the nonlinear solves.

\LL{Note that the return mapping algorithm for the Drucker--Prager model in Section~\ref{sec:mech} admits a closed-form solution, so its evaluation requires no local iterations and remains computationally inexpensive per quadrature point.} For more complex constitutive models, where the return mapping requires solving a nonlinear problem at each quadrature point, the full-order assembly becomes more expensive and the gains of the hyper-reduction are expected to be larger.

\begin{table}[htbp]
\centering
\caption{Wall-clock times, in seconds, for three test simulations spanning the plastification range, on the three meshes. Each time is accumulated over the 312 steps of the transient simulation. The total includes minor overheads not itemized in the breakdown. For the RB model, the solve column includes the Galerkin projection of the assembled system onto the reduced basis. For the HR-RB model, the online precomputation of the reduced baseline operators (precomp.) is included in the total, so the speed-ups compare full online costs. \AM{The RB model decreases only the solve time, and the assembly and the return mapping still scale with the mesh. The HR-RB model also cuts these two costs, so its speed-up grows with the refinement and with the plastification.}\label{tab:ex2d-timing}}
\begin{tabular}{llcccccc}
\toprule
& model & return map & assembly & solve & precomp. & total & speed-up \\
\midrule
\multicolumn{8}{l}{Mesh $990$ elements}\\
$3\%$ plastic  & FOM   & 2.0 & 12.6 & 8.4  & --  & 25.0 & $1\times$ \\
               & RB    & 2.3 & 13.8 & 2.4  & --  & 20.8 & $1.2\times$ \\
               & HR-RB & 0.8 & 1.6  & 0.6  & 0.3 & 3.5  & $7.2\times$ \\[2pt]
$42\%$ plastic & FOM   & 3.1 & 14.6 & 10.3 & --  & 30.6 & $1\times$ \\
               & RB    & 3.4 & 15.4 & 2.8  & --  & 24.3 & $1.3\times$ \\
               & HR-RB & 1.0 & 1.8  & 0.7  & 0.3 & 4.0  & $7.6\times$ \\[2pt]
$76\%$ plastic & FOM   & 5.0 & 17.3 & 12.8 & --  & 38.3 & $1\times$ \\
               & RB    & 4.9 & 16.6 & 3.0  & --  & 27.5 & $1.4\times$ \\
               & HR-RB & 1.1 & 1.9  & 0.7  & 0.3 & 4.3  & $8.9\times$ \\
\midrule
\multicolumn{8}{l}{Mesh $2\,963$ elements}\\
$3\%$ plastic  & FOM   & 5.8  & 36.8 & 15.5 & --  & 63.5  & $1\times$ \\
               & RB    & 6.1  & 40.5 & 7.0  & --  & 59.9  & $1.1\times$ \\
               & HR-RB & 0.8  & 1.8  & 0.8  & 0.8 & 4.5   & $14.2\times$ \\[2pt]
$42\%$ plastic & FOM   & 9.2  & 44.1 & 20.4 & --  & 80.6  & $1\times$ \\
               & RB    & 9.4  & 45.3 & 8.2  & --  & 70.1  & $1.2\times$ \\
               & HR-RB & 1.1  & 2.0  & 0.8  & 0.8 & 5.1   & $16.0\times$ \\[2pt]
$76\%$ plastic & FOM   & 15.1 & 52.7 & 25.7 & --  & 102.3 & $1\times$ \\
               & RB    & 14.2 & 48.8 & 8.9  & --  & 79.8  & $1.3\times$ \\
               & HR-RB & 1.2  & 2.1  & 0.9  & 0.8 & 5.4   & $19.1\times$ \\
\midrule
\multicolumn{8}{l}{Mesh $9\,982$ elements}\\
$3\%$ plastic  & FOM   & 18.9 & 122.4 & 41.1 & --  & 199.8 & $1\times$ \\
               & RB    & 21.1 & 135.9 & 23.3 & --  & 200.0 & $1.0\times$ \\
               & HR-RB & 0.9  & 1.9   & 0.9  & 2.4 & 6.7   & $29.7\times$ \\[2pt]
$42\%$ plastic & FOM   & 31.8 & 153.0 & 56.0 & --  & 264.3 & $1\times$ \\
               & RB    & 31.6 & 149.8 & 26.8 & --  & 230.8 & $1.1\times$ \\
               & HR-RB & 1.1  & 2.2   & 1.0  & 2.4 & 7.3   & $36.4\times$ \\[2pt]
$76\%$ plastic & FOM   & 50.8 & 174.6 & 67.2 & --  & 320.5 & $1\times$ \\
               & RB    & 48.8 & 162.5 & 29.9 & --  & 266.5 & $1.2\times$ \\
               & HR-RB & 1.1  & 2.2   & 1.0  & 2.4 & 7.4   & $43.5\times$ \\
\bottomrule
\end{tabular}
\end{table}

\subsection{Three-dimensional dam}\label{sec:ex3d}

\subsubsection{Problem setup}\label{sec:ex3d-setup}

\begin{figure}[htbp]
\centering
\begin{minipage}[t]{.8\linewidth}
\centering
\includegraphics[width=\linewidth]{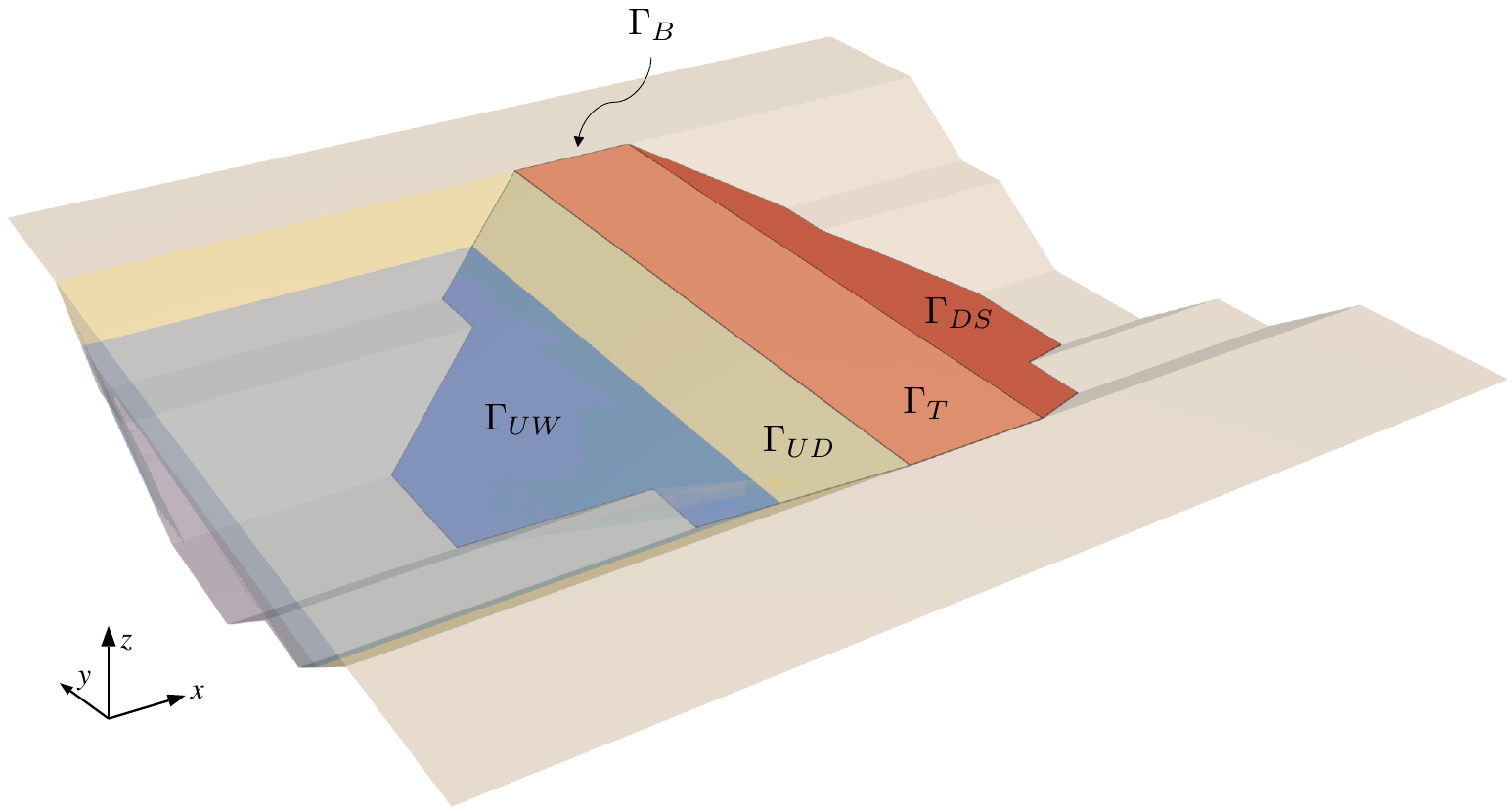}\\
(a)
\end{minipage}\\[8pt]
\begin{minipage}[t]{.43\linewidth}
\centering
\includegraphics[width=\linewidth]{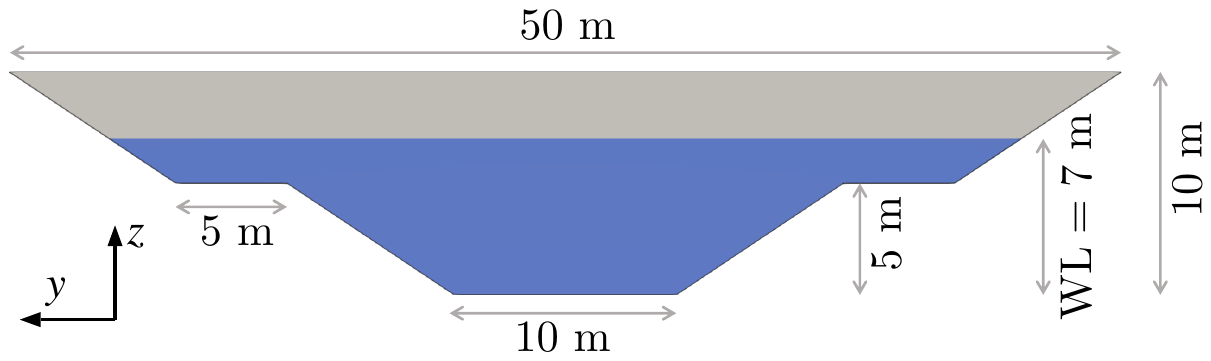}\\
(b)
\end{minipage}\hfill
\begin{minipage}[t]{.46\linewidth}
\centering
\includegraphics[width=\linewidth]{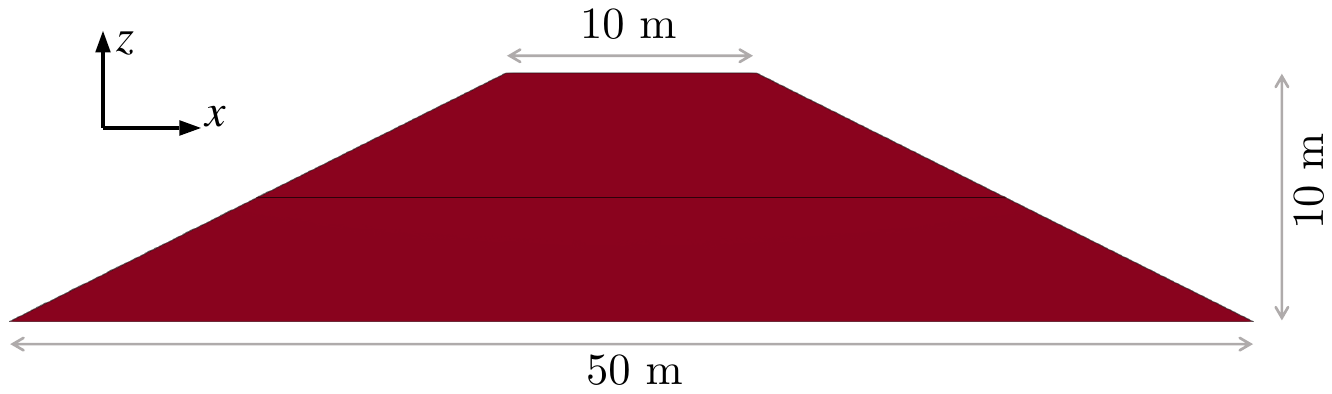}\\
(c)
\end{minipage}\\[8pt]
\begin{minipage}[t]{.33\linewidth}
\centering
\includegraphics[width=\linewidth]{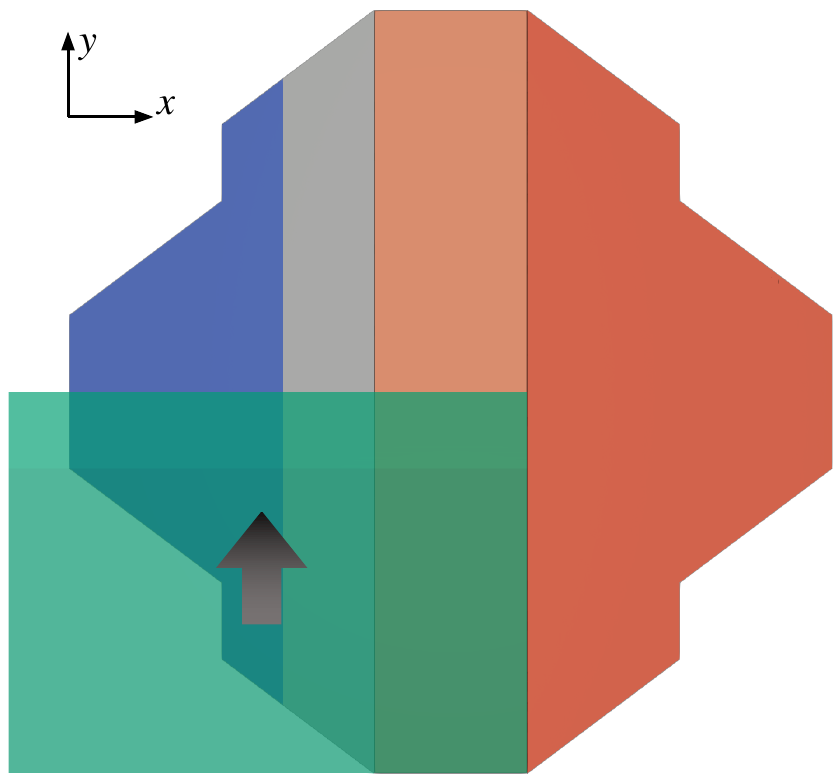}\\
(d)
\end{minipage}
\caption{The 3D valley dam. 
	(a) Perspective view. The computational domain is the dam body, with the boundary partitioned into the wet and dry upstream faces $\Gamma_{UW}$ and $\Gamma_{UD}$, the crest $\Gamma_T$, the downstream face $\Gamma_{DS}$, and the contact with the valley $\Gamma_B$ (base, benches and valley walls). The valley terrain, the retained tailings and the water are not discretized, and act on the dam through the boundary conditions. 
	(b) Upstream elevation, with the dimensions along the dam axis. 
	(c) Cross-section, with the cross-valley dimensions. 
	(d) Plan view, with the advancing deposition layer and the direction of its front.}
\label{fig:ex3d-domain}
\end{figure}

The dam sits in a V-shaped valley, with the geometry and dimensions of Figure~\ref{fig:ex3d-domain}. In the coordinates used below, $x$ runs across the valley, $y$ along the dam axis and $z$ is the elevation. 
The cross-section at the middle of the valley is the trapezoid of the two-dimensional example in Section \ref{sec:ex2d}. 
The valley walls open as the elevation increases, with a horizontal bench at mid-height, and the crest is $L=50$ m long while the base is only $10$ m. 
The retained tailings again reach the crest, and the water level splits the upstream face into its wet and dry parts, $\Gamma_{UW}$ and $\Gamma_{UD}$.

As in two dimensions, we simulate a raise of the dam of height
$h_{\mathrm{fill}}$ and the subsequent consolidation, with the parameters
of Table~\ref{tab:ex2d-params}.
The boundary conditions extend those of Section~\ref{sec:ex2d-setup}. 
On the part of the boundary that is in contact with the valley, $\Gamma_R^u=\Gamma_B$, we impose a Winkler foundation \eqref{eq:bc-mech}, with modulus $k_n=10^{9}\,\mathrm{Pa/m}$ in the normal direction and $k_t=10^{6}\,\mathrm{Pa/m}$ in the tangential one.
The dam is therefore supported by the valley but not rigidly fixed to it, so that it is able to settle under the raise of the dam.
On the upstream faces, we apply the tractions of the two-dimensional example \eqref{eq:ex2d-tractions}, with the elevation $z$ in place of $y$ and the rise load magnitude renamed $q_z$,
\begin{equation*}
  \hat{\mathbf t}=\big(q_x+K_a\,\sigma_v',\;0,\;-q_z-\sigma_v'\big)^{\!\top}
  \text{ on }\Gamma_{UD},
  \qquad
  \hat{\mathbf t}=\big(q_x+K_a\,\sigma_v',\;0,\;-q_z-\sigma_v'\big)^{\!\top}
  -p_w\,\bn
  \text{ on }\Gamma_{UW},
\end{equation*}
so the loads act across the valley and vertically, with no component along the dam axis. The raise loads the crest, with $\hat{\mathbf t}=(0,0,-q_z)^{\!\top}$ on $\Gamma_T$, and the upstream faces through the tailings, by the terms $q_x$ and $q_z$ above. The downstream face $\Gamma_{DS}$ is traction-free.
For the flow problem, the wet upstream face is a Dirichlet boundary with the hydrostatic pressure, $\hat p=\gw(\mathrm{WL}-z)$ on $\Gamma_D^p=\Gamma_{UW}$. The downstream face is a seepage face with the same seepage length $\delta$, and all the remaining boundaries, including the contact with the valley, are impervious.

\begin{figure}[htbp]
\centering
\includegraphics[width=\linewidth]{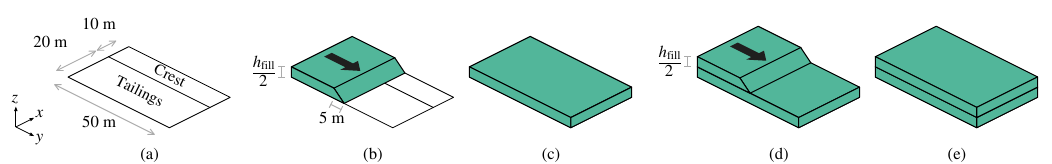}
\caption{Staged raise on the three-dimensional dam.
	(a) Deposition surface, formed by the crest and the strip of tailings behind it.
	(b) The first layer is deposited along the dam axis, in the direction of the arrow.
	(c) First layer complete, at day $5$.
	(d) Deposition of the second layer.
	(e) Both layers complete, at day $10$, so that the raise reaches its full height $h_{\mathrm{fill}}$.
	The thickness of the layers is exaggerated for visualization purposes.}
\label{fig:ex3d-fill}
\end{figure}

In this case, the raise is not applied as a uniform ramp in time, see \eqref{eq:ex2d-ramp}. Following Nasika et al. \cite{Nasika2023}, the new material is placed in two layers of height $h_{\mathrm{fill}}/2$, deposited by a front that travels along the dam axis at constant speed, in two sweeps of
$5$ days each, as sketched in Figure~\ref{fig:ex3d-fill}. The deposition front is inclined, that is, the layer thickness increases linearly over a length of $5\,\mathrm{m}$ until the layer reaches its full height $h_{\mathrm{fill}}/2$. The vertical load $q_z(y,t)$ is the weight of the material deposited at each position $y$ along the axis, with $q_x=K_a\,q_z$ as in two dimensions. Each position along the axis therefore supports the full raise behind both fronts, half of it between them and none ahead of them. After day 10 both layers are complete and the load reaches the value $\gamma_f\,h_{\mathrm{fill}}$ of the two-dimensional ramp \eqref{eq:ex2d-ramp}.
The inclination smooths the loading as the front passes each point, which keeps the elastoplastic response stable numerically.

\begin{comment}
The load \eqref{eq:ex2d-ramp} becomes
\begin{equation}
  q_z(y,t)=\bigl(\chi_1(y,t)+\chi_2(y,t)\bigr)\,\gamma_f\,h_{\mathrm{fill}},
  \qquad
  q_x=K_a\,q_z,
  \label{eq:ex3d-load}
\end{equation}
where $\chi_k$ is the fraction of the raise placed by sweep $k$ at the
position $y$ along the axis, and $f_k(t)$ is the position of its front.
Ahead of the front there is no new material; behind it, the layer grows
linearly over a beach length $w=5\,\mathrm{m}$ and then has its full
height,
\begin{equation}
  \chi_k(y,t)=
  \begin{cases}
    0, & y\ge f_k(t),\\[2pt]
    \frac{f_k(t)-y}{2w}, & f_k(t)-w\le y\le f_k(t),\\[2pt]
    \frac{1}{2}, & y\le f_k(t)-w.
  \end{cases}
  \label{eq:ex3d-front}
\end{equation}
A crest point therefore supports the full raise behind both fronts, half
of it between them and none ahead of them, and after day 10 both layers
are complete.
\end{comment}

The initial condition is the steady state before the raise, and the transient covers the same 30 days on the time grid of the two-dimensional example. 
The domain is discretized with a mesh of $117\,883$ tetrahedra (Table~\ref{tab:ex3d-mesh}). With Taylor--Hood $P_2$--$P_1$ elements, it has $N=551\,116$ unknowns, twelve times the finest 2D discretization in Section \ref{sec:ex2d}. Element integrals are also computed with a quadrature rule of order five, which now has fifteen points per tetrahedron.

\begin{table}[htbp]
\centering
\caption{The 3D mesh: number of elements, of quadrature points, of displacement and pressure degrees of freedom, and total number of unknowns.\label{tab:ex3d-mesh}}
\begin{tabular}{lccccc}
\toprule
mesh & $|\mathcal E|$ & quad.\ points & $n_u$ & $n_p$ & $N = n_u + n_p$ \\
\midrule
3D & 117\,883 & 1\,768\,245 & 527\,169 & 23\,947 & 551\,116 \\
\bottomrule
\end{tabular}
\end{table}

The parameters, the sampling and the validation protocol are those of Section~\ref{sec:ex2d-train}. The fraction of quadrature points plastified at the end of the simulation now ranges from $5\%$ to $39\%$ over the training set, a narrower span than in two dimensions, since the valley walls support part of the load and the dam yields less than in plane strain.

\subsubsection{Reduced-order results}\label{sec:ex3d-results}

The POD bases are computed with the tolerance $\varepsilon_{\mathrm{POD}}=10^{-3}$ in the criterion \eqref{eq:podtrunc}, giving $k_u=34$ and $k_p=76$ modes. The reduced problem has 110 unknowns, in place of the $551\,116$ of the full-order model.

\begin{figure}[htbp]
\centering
\begin{minipage}[c]{0.59\linewidth}
\centering
\includegraphics[width=\linewidth]{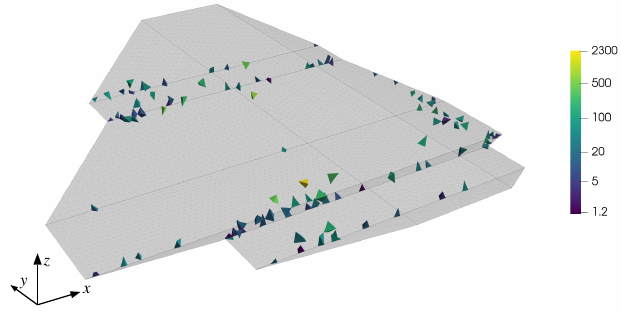}
\end{minipage}\hfill
\begin{minipage}[c]{0.38\linewidth}
\centering
\includegraphics[width=\linewidth]{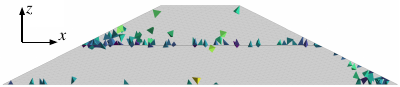}\\[10pt]
\includegraphics[width=.6\linewidth]{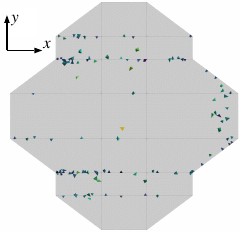}
\end{minipage}
\caption{\LL{Mechanical} ECSW rule $\tilde{\mathcal E}_u$ of the
three-dimensional dam (110 elements), coloured by the weights $\xi^e$ on a logarithmic colour scale, with the mesh in transparent grey: perspective view (left), and elevation and plan views (right). The elements spread over the dam body, with concentrations along the contact with the valley walls and the base.}\label{fig:ex3d-rules-u}

\centering
\begin{minipage}[c]{0.59\linewidth}
\centering
\includegraphics[width=\linewidth]{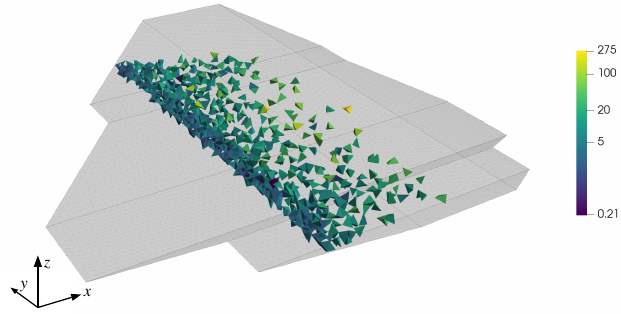}
\end{minipage}\hfill
\begin{minipage}[c]{0.38\linewidth}
\centering
\includegraphics[width=\linewidth]{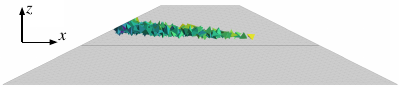}\\[10pt]
\includegraphics[width=.6\linewidth]{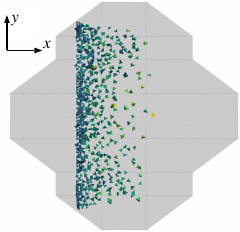}
\end{minipage}
\caption{Flow ECSW rule $\tilde{\mathcal E}_p$ of the three-dimensional dam (611 elements), coloured by the weights $\xi^e$ on a logarithmic colour scale, with the mesh in transparent grey: perspective view (left), and elevation and plan views (right). The selection concentrates in the narrow band along the phreatic surface.}\label{fig:ex3d-rules-p}
\end{figure}

The decomposed ECSW rules, trained with the tolerance $\tau=10^{-2}$ in \eqref{eq:nnls}, select $110$ of the $117\,883$ elements for the \LL{mechanical} block ($0.09\%$) and $611$ for the flow one ($0.52\%$).
Figures~\ref{fig:ex3d-rules-u} and~\ref{fig:ex3d-rules-p} show the two reduced meshes. \LL{Similarly to the two-dimensional case, the selected elements concentrate in the regions exhibiting strong nonlinearities.} The \LL{elements of $\tilde{\mathcal E}_u$} spread over the dam body, with concentrations along the contact with the valley walls and the base, and \LL{those of $\tilde{\mathcal E}_p$} gather around the phreatic \LL{surface}.

\begin{figure}[htbp]
\centering
\includegraphics[width=.95\linewidth]{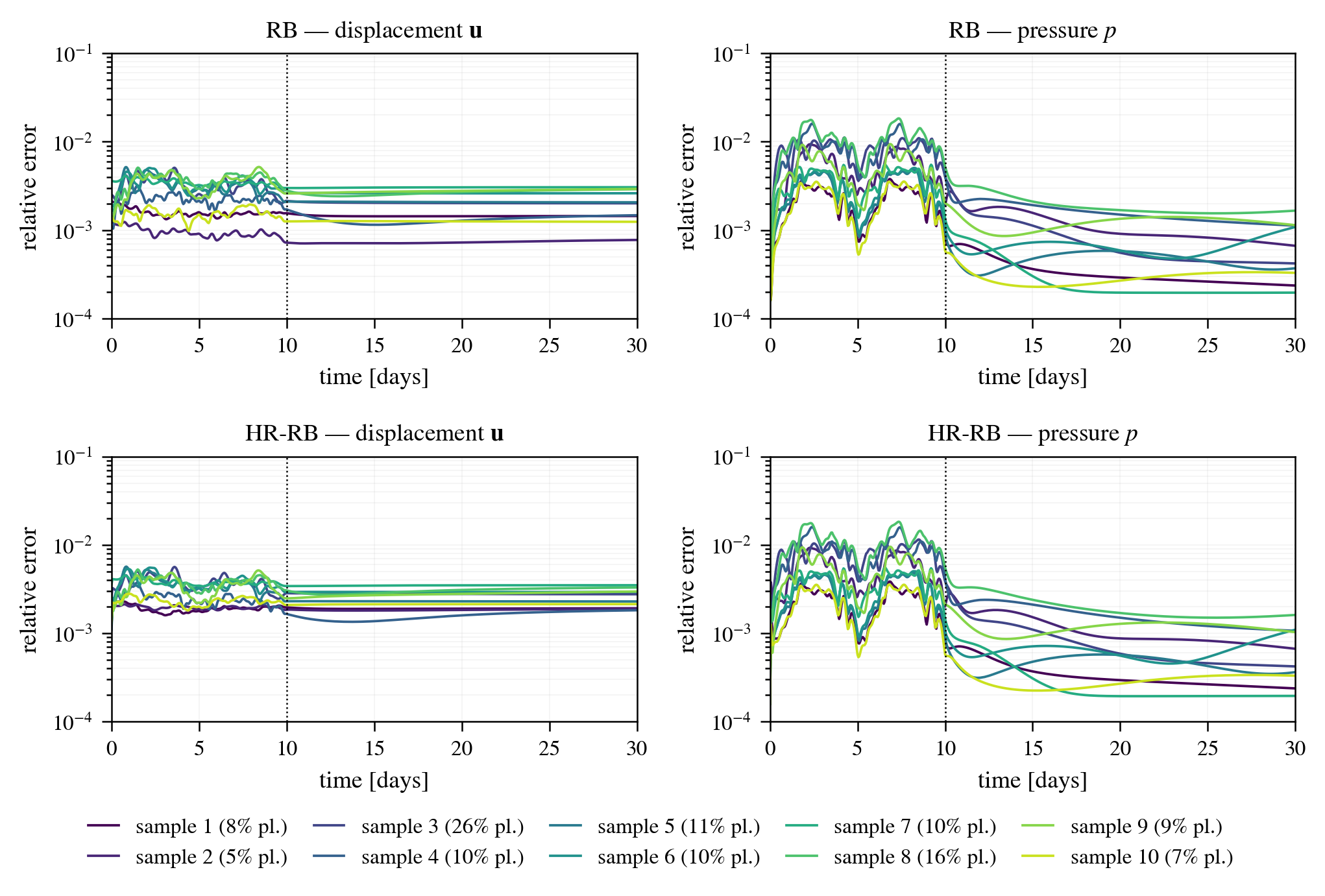}
\caption{Evolution in time of the per-step relative errors on the ten test samples of the three-dimensional dam, for the RB (top row)
and the HR-RB (bottom row) models, with the displacement on the left and the pressure on the right. The samples correspond to the parameter points of Figure~\ref{fig:ex2d-err}, and the legend indicates the fraction of quadrature points plastified at the end of each simulation. The dotted line marks the end of the deposition at day 10. \AM{The errors peak while the deposition fronts travel and decay during the consolidation, staying below $1.9\%$ in all cases.}}\label{fig:ex3d-err}
\end{figure}

The reduced models are validated on the ten test simulations.
Figure~\ref{fig:ex3d-err} shows the evolution in time of the per-step relative errors, and Table~\ref{tab:ex3d-errors} collects the mean global and maximum errors. Both reduced models reproduce the displacement to about $0.3\%$ and the pressure to about $0.5\%$ in the global error \eqref{eq:ex2d-err}. The per-step errors peak while the deposition fronts travel and decay during the consolidation that follows, staying below $1.9\%$ in all cases, which is acceptable for this application.
The HR-RB errors sit on the error floor of the RB model, so the ECSW cubature adds no visible error.

\begin{table}[htbp]
\centering
\caption{Mean relative errors over the 10 test simulations --- global
errors \eqref{eq:ex2d-err} and maximum errors --- for the RB and HR-RB
models on the three-dimensional dam. \AM{The HR-RB errors sit on the error floor of the RB model, so the cubature adds no visible error.}\label{tab:ex3d-errors}}
\begin{tabular}{lcccc}
\toprule
model & $e_u$ & $e_p$ & $e_u^{\mathrm{max}}$ & $e_p^{\mathrm{max}}$ \\
\midrule
RB    & 2.39e-3 & 4.58e-3 & 3.81e-3 & 8.71e-3 \\
HR-RB & 2.80e-3 & 4.62e-3 & 4.28e-3 & 8.75e-3 \\
\bottomrule
\end{tabular}
\end{table}

Table~\ref{tab:ex3d-timing} reports wall-clock times for three test
simulations spanning the plastification range of the test set. The
full-order simulations take $11$ to $14$ hours. \AM{As in two dimensions, the RB model is only able to reduce the solve time, and takes about $6$ hours. In this case, the slight decrease in its assembly and return-mapping times is due to the method taking between $18\%$ and $27\%$ fewer Newton--Raphson iterations than the full-order model. The HR-RB model, in contrast, cuts the assembly and the return mapping themselves, and runs the simulations in $10$ to $11$ minutes, including the precomputation of the reduced baseline operators.} The speed-ups are $65$--$77\times$, against the
$2.0$--$2.2\times$ of the RB model. As observed in two dimensions, the acceleration grows with the plastification of the simulation. In this case, the return-mapping time collapses from about $50$ minutes to less than $5$ seconds.

These gains are larger than any observed in two dimensions, and they
follow the trend of Section~\ref{sec:ex2d-results}. The cost of the
full-order model is set by the discretization, and grows with the size of the mesh. The cost of the hyper-reduced model is set by the physics, through the reduced dimensions and the size of the rules, which grow far more slowly. The benefit of the hyper-reduction is therefore largest where the full-order model is most expensive. The discretization can thus be chosen on accuracy grounds alone, and the hyper-reduction confines the expensive nonlinear evaluations to the regions that require them.

\begin{table}[htbp]
\centering
\caption{Wall-clock times, in minutes, for three test simulations of the
three-dimensional dam, spanning the plastification range of the test set
(samples 2, 8 and 3 of Figure~\ref{fig:ex3d-err}). Each time is
accumulated over the 312 steps of the transient simulation; columns as in
Table~\ref{tab:ex2d-timing}. \AM{The pattern of the two-dimensional example is amplified on this larger model, where the assembly and the return mapping dominate the full-order cost and the hyper-reduction reduces them significantly.}\label{tab:ex3d-timing}}
\begin{tabular}{llcccccc}
\toprule
& model & return map & assembly & solve & precomp. & total & speed-up \\
\midrule
$5\%$ plastic  & FOM   & 40.8 & 304.1 & 329.3 & --  & 680.1 & $1\times$ \\
               & RB    & 35.4 & 260.4 & 44.4  & --  & 345.2 & $2.0\times$ \\
               & HR-RB & 0.07 & 3.0   & 2.9   & 4.2 & 10.5  & $64.8\times$ \\[2pt]
$16\%$ plastic & FOM   & 48.6 & 348.4 & 391.4 & --  & 795.4 & $1\times$ \\
               & RB    & 39.8 & 281.8 & 49.4  & --  & 376.5 & $2.1\times$ \\
               & HR-RB & 0.08 & 3.3   & 3.2   & 4.2 & 11.2  & $71.3\times$ \\[2pt]
$26\%$ plastic & FOM   & 50.0 & 361.4 & 403.1 & --  & 821.8 & $1\times$ \\
               & RB    & 37.0 & 279.3 & 48.2  & --  & 369.8 & $2.2\times$ \\
               & HR-RB & 0.07 & 3.1   & 3.0   & 4.2 & 10.7  & $76.7\times$ \\
\bottomrule
\end{tabular}
\end{table}

\section{Conclusions}\label{sec:concl}

This paper presents a hyper-reduced order model for coupled elastoplastic hydro-mechanical problems in partially saturated soils. 
The displacement and the pore pressure are approximated in a field-split POD basis, and the Galerkin-projected equations are integrated by decomposed ECSW cubatures, one per \LL{balance equation}. 
Each cubature is trained on the correction of its residual with respect to an exact affine baseline, an
elastic response for the \LL{mechanical} block and a frozen-coefficient state for the flow one, whose reduced form is precomputed and adds no significant cost to the online stage. 
The construction uses the same training snapshots as the POD bases, never re-runs the full-order model, is independent of the constitutive law, and is implemented in FEniCSx through the assembly of the
original variational forms, restricted to a subdomain.

The strategy is tested on two dams retaining tailings and water, simulating a raise of the dam and the subsequent consolidation \LL{phase}. The ROM uses four parameters, the friction angle $\phi$, the Young's modulus $E$ and the saturated conductivity $K_s$ of the dam and the height of the raise $h_{\mathrm{fill}}$, which span a wide range of responses, \LL{from nearly elastic behaviour throughout the domain to extensive plastic zones covering most of the dam}.
On the two-dimensional dam, the HR-RB model reproduces the displacement and pressure fields to about $0.1\%$ of the full-order solution on the test simulations. The errors sit on the error floor of the RB model, so the accuracy is limited by the basis truncation and not by the cubature.
The speed-ups grow from $7$--$9\times$ on the coarsest mesh to $30$--$43\times$ on the finest one. 
The reduced dimensions and the reduced meshes track the plastified zones and the phreatic \LL{surface}, whose extent is set by the physics and not by the resolution, while the full-order cost keeps growing with the mesh. The hyper-reduction can thus be regarded as an adaptive strategy in space. The mesh can be uniform and as fine as needed to resolve the physics \LL{under consideration}, without economizing on the number of elements, and the selection automatically restricts the nonlinear work to the regions where it matters.

The three-dimensional dam confirms this behaviour on a full-order model an order of magnitude larger, with a deposition process that makes the response fully three-dimensional.
The errors stay at about $0.3\%$ in displacement and $0.5\%$ in pressure, and the online speed-ups reach $65$--$77\times$, turning simulations of $11$ to $14$ hours into runs of $10$ to $11$ minutes.

\LL{Finally, we emphasize that the constitutive model choices are independent of the hyper-reduction strategy. The proposed methodology does not exploit the specific mathematical form of the mechanical constitutive model, the water retention curve, or the relative permeability model. The employed associated Drucker--Prager model, the van Genuchten model and the cubic dependence of the hydraulic conductivity on the saturation act on the reduced-order model solely through the assembled residuals. Consequently, the formulation extends analogously to any other mechanical constitutive law, or any other constitutive choices for the hydraulic subproblem, without any modification of the construction of the bases, the cubatures or the selection.}

%========================================================================
\section*{Acknowledgments}

This work received funding from the Spanish Ministry of Science and Innovation, AEI/\allowbreak 10.13039/\allowbreak 501100011033, through Grant numbers PID2023\allowbreak-148952OB-I00 and PID2023\allowbreak-153082OB-I00, from the HE-MSCA-DN EarthSafe (G.A. n 101120556), and from the Generalitat de Catalunya, 2021\allowbreak-SGR\allowbreak-01049.
AM is Fellow of the Serra H\'unter Programme of the Generalitat de Catalunya.

\section*{Data Availability Statement}

The data that support the findings of this study are available from the corresponding author upon reasonable request.

%========================================================================

\appendix

\section{Blocks of the Jacobian}\label{app:jacobian}

At each time step, problem \eqref{eq:wf-m}--\eqref{eq:wf-f} is solved for the displacement $\bu^{n}$ and the pressure $p^{n}$ at time $t^{n}$, the state at $t^{n-1}$ being known. The blocks of the Jacobian in \eqref{eq:newton} follow from differentiating the residuals with respect to the two unknowns. They are written below as bilinear forms in the increments $(\delta\bu,\delta p)\in\mathcal V_0\times\mathcal W_0$ and the test functions $(\bv,w)\in\mathcal V_0\times\mathcal W_0$.
All the coefficients are evaluated at the current Newton--Raphson iterate, and a prime denotes their derivative
with respect to the pressure, $(\cdot)'=\partial(\cdot)/\partial p$,
\begin{align}
  J_{uu}(\delta\bu,\bv) &=
    \int_\Omega \beps(\bv)\!:\!\mathbb C_{\mathrm{tan}}\!:\!\beps(\delta\bu)\,\mathrm{d}\Omega
    +\int_{\Gamma_R^u}\!\big(k_n\,(\delta\bu\!\cdot\!\bn)(\bv\!\cdot\!\bn)
    +k_t\,\delta\bu_t\!\cdot\!\bv_t\big)\,\mathrm{d}\Gamma,
    \label{eq:Juu}\\
  J_{up}(\delta p,\bv) &=
    -\int_\Omega \big(S_e+S_e'\,p^{n}\big)\,\delta p\,\boldsymbol\nabla\!\cdot\!\bv\,\mathrm{d}\Omega
    -\int_\Omega \rho'\,\delta p\,\bg\!\cdot\!\bv\,\mathrm{d}\Omega,
    \label{eq:Jup}\\
  J_{pu}(\delta\bu,w) &=
    \int_\Omega \Theta\,\boldsymbol\nabla\!\cdot\!(\delta\bu)\,w\,\mathrm{d}\Omega,
    \label{eq:Jpu}\\
  J_{pp}(\delta p,w) &=
    \Delta t\!\int_\Omega \tfrac{K}{\gw}\,\boldsymbol\nabla\delta p\!\cdot\!\boldsymbol\nabla w\,\mathrm{d}\Omega
    +\Delta t\!\int_\Omega \tfrac{K'}{\gw}\,\delta p\,\boldsymbol\nabla p^{n}\!\cdot\!\boldsymbol\nabla w\,\mathrm{d}\Omega
    \nonumber\\
  &\quad
    +\int_\Omega \Big(C+\tfrac{\Theta}{K_w}\Big)\,\delta p\,w\,\mathrm{d}\Omega
    +\int_\Omega \Big(C'+\tfrac{\Theta'}{K_w}\Big)\,\delta p\,(p^{n}-p^{n-1})\,w\,\mathrm{d}\Omega
    \nonumber\\
  &\quad
    +\int_\Omega \Theta'\,\delta p\,\boldsymbol\nabla\!\cdot\!(\bu^{n}-\bu^{n-1})\,w\,\mathrm{d}\Omega
    -\Delta t\!\int_\Omega \tfrac{K'}{\gw}\,\delta p\,\rho_w\bg\!\cdot\!\boldsymbol\nabla w\,\mathrm{d}\Omega
    \nonumber\\
  &\quad
    +\Delta t\!\int_{\Gamma_R^p}\!\beta\,\mathcal{H}(p^{n})\,\delta p\,w\,\mathrm{d}\Gamma,
    \label{eq:Jpp}
\end{align}
where $\mathcal H$ is the Heaviside step function, smoothed as in Section~\ref{sec:weak}, and which restricts the seepage term to the part of $\Gamma_R^p$ where the boundary is saturated ($p^{n}>0$), and $\mathbb C_{\mathrm{tan}}=\partial\bsig'/\partial\beps$ is the consistent tangent of the return mapping \cite{deSouzaNeto2008}.

\bibliographystyle{unsrtnat}
\bibliography{references}

\end{document}